\documentclass[preprint,showpacs,showkeys,preprintnumbers,amsmath,amssymb,nofootinbib
]{revtex4}
\usepackage{lscape}
\usepackage{graphicx}
\usepackage{dcolumn}
\usepackage{bm}
\usepackage{mathtools}
\usepackage{fancybox}
\usepackage[dvipsnames,usenames]{color}
\def\beq{\begin{equation}}
\def\eeq{\end{equation}}
\def\beqna{\begin{eqnarray}}
\def\eeqna{\end{eqnarray}}
\def\eq(#1){Eq.(\ref{#1})}
\def\Eq(#1){Equation (\ref{#1})}
\def\eqs(#1)-(#2){Eqs.(\ref{#1}) and (\ref{#2})}
\def\Eqs(#1)-(#2){Equations (\ref{#1}) and (\ref{#2})}
\def\eqss(#1)-(#2){Eqs.(\ref{#1})-(\ref{#2})}
\def\Eqss(#1)-(#2){Equations (\ref{#1})-(\ref{#2})}
\def\fig#1{FIG.\ref{#1}}
\def\figs#1-#2{FIGs.\ref{#1} and \ref{#2}}

\def\grad{\nabla}

\def\d#1{\!\!{\rm d}#1}
\def\dn#1#2{\!\!{\rm d}^{#1}{#2}\,}

\def\del#1{\partial_{#1}}
\def\v#1{{\bf #1}}
\def\h#1{\hat{#1}}
\def\t#1{\tilde{#1}}
\def\hv#1{\hat{{\bf #1}}}

\def\c#1{{\cal #1}}

\def\evac{\epsilon_{0}}

\allowdisplaybreaks[2]
\begin{document}
\preprint{APS/123-QED}
%
\title{Theory of Single Susceptibility for Near-field Optics\\
Equally Associated with Scalar and Vector Potentials
}
\author{Itsuki Banno}
\email{banno@yamanashi.ac.jp}
\affiliation{
Interdisciplinary Graduate School of Medicine and Engineering,
University of Yamanashi,
4-3-11 Takeda, Kofu, Yamanashi 400-8511, Japan }
\date{\today}
\begin{abstract}
A nonlocal response theory was developed  
to describe a many-electron system 
within the neighborhood of a nanostructure
radiating  the longitudinal and transverse electric fields,
which are
fundamentally reduced to the scalar and vector potentials (SP and VP).
The coexistence of the SP and VP incidences distinguishes such a near-field optical
system from the ordinary optical system, in which only the VP (under the Coulomb gauge)
incidence survives far from the light source. 
This fact is the motivation for equal treatment of 
the SP and VP as the cause of the response in the near-field optical system.
In the semiclassical treatment, 
the linear and nonlinear single susceptibilities  
are derived in the form of Heisenberg operators 
by the functional derivatives of the action integral of the matter 
with respect to the SP and VP.
These single susceptibilities relate the SP and VP (as the cause) to the induced charge and current densities (as the result), and guarantee charge conservation and gauge invariance;
this theory is free from gauge-fixing.
It is necessary to consider the quantum many-electron effect (exchange-correlation effect) 
to make the ground state bounded in the non-perturbed system.
This is done by employing the fundamental idea of density functional theory,
instead of the ordinary unequal treatment of the SP and VP, 
that is, remaking the SP into a Coulomb interaction between electron charges. 
Applying the present linear response theory to the non-metallic material in a limited near-field optical system
reveals that the electric field with the associated permittivity is not suitable quantity to 
describe the response,
 instead, the SP and VP with associate single susceptibility are essential.
\end{abstract}
\maketitle
%
\section{Introduction}
\label{sec:introduction}
%
This paper develops a nonlocal response theory adequate for
near-field optics (NFO) in the semiclassical treatment. 
The linear and nonlinear single susceptibilities are derived systematically
by the functional derivatives of the action integral of the matter with respect 
to the scalar and vector potentials (SP and VP).
These linear and nonlinear single susceptibilities relate
the SP and VP (as the cause) to the induced charge and current densities (as the result), 
and guarantee charge conservation and gauge invariance.   
The present single susceptibilities and associated induced charge and current densities
are given in the form of Heisenberg operators.

In Ref.\cite{SpringerFallacy}, the present author discussed 
the linear single susceptibility, its application to an one-electron optical system, and
a naive idea of employing the density functional theory.
This  paper is its generalization including systematic derivation of 
linear and nonlinear single susceptibilities in the form of Heisenberg operator,
a simple proof of charge conservation and gauge invariance guaranteed by 
such the susceptibilities, and application to a many-electron system 
with detailed discussion on the density functional theory.
 
The introduction below contains the followings:
\S\ref{sec:singleSus} reveals the necessity of the single susceptibility, instead of 
the electric permittivity and magnetic permeability.
\S\ref{sec:equalTreatment} points out the preference to equal treatment of the SP and VP 
as the cause of response in NFO, 
instead of the unequal treatment in ordinary optics under the Coulomb gauge.
\S\ref{sec:MEP} explains the difficulty of constructing the response theory in NFO, 
which inevitably connected to a many-electron problem via the SP.
\S\ref{sec:purpose} represents the purpose of this paper .
\subsection{The necessity of the single susceptibility}
\label{sec:singleSus}
As the cause of response, it is natural and essential to use the SP and VP, 
which represent for the electromagnetic (EM) field in the Hamiltonian for quantum electrodynamics.
Three reasons are given below for the 
{\it inapplicability of the electric and magnetic fields} as the cause of
response.
First, there exist such systems that cannot be described in terms of the electric and/or
magnetic fields, namely, the superconductor system with the Meissner effect\cite{London} and 
the coherent electron system with the Aharanov-Bohm effect\cite{ABeffect}. 
A limited NF optical system is another example, 
as shown in the one-electron system in Ref.\cite{SpringerFallacy} 
(and will be shown in a many-electron system in \S\ref{sec:oneElectron} of this paper).

Second, the constitutive equations with 
the electric permittivity and magnetic permeability 
give relationships between redundant degrees of freedom. 
Actually, the essential source of the EM field is the three components of charge density and the transverse current density. 
The longitudinal current density is excluded because it can be determined through charge conservation law, once the charge density is known.
However, the polarization and magnetization as the source of the EM field 
have totally six components, which include the redundancy. 
So that the associated constitutive equations using the two susceptibilities 
include the constraint condition for the redundancy, of which the physical meaning is not declared.
This situation is physically unreasonable and should be fixed by
the constitutive equation using a single susceptibility associated with the proper degrees of freedom.

Third, as first claimed by Cho\cite{Cho01,Cho02}
for the low-symmetry systems with chirality (such as the NF optical system 
with a skewed nanostructure),
the ordinary two constitutive equations are not available because 
the electric and magnetic responses become indistinguishable.
He also revealed that this error cannot be fixed by the Drude-Born-Fedorov 
fomulas\cite{DBF}, which extends the two constitutive equations adding the cross terms of the electric-field-induced magnetization and the magnetic-field-induced polarization. 

Therefore,  from a general view point, it is essential to employ 
a single susceptibility with the SP and VP, 
instead of the electric permittivity and magnetic permeability 
with the electric and magnetic fields.
%
\subsection{The preference to equal treatment of the SP and VP in a NF optical system}
\label{sec:equalTreatment}
%
\begin{figure}[b]
\begin{center}
\includegraphics[width=0.5\textwidth]{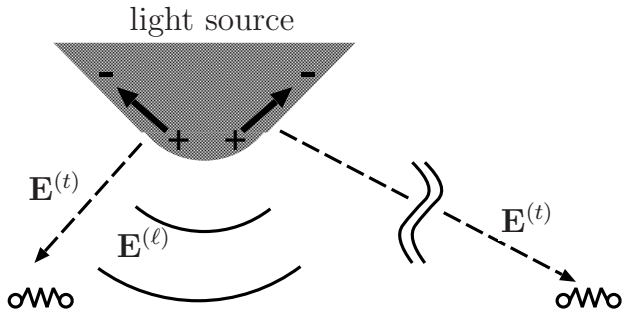}
\caption{
Optical systems under near- and far-field incidences (the left and right side figures, respectively) .
The former system is exposed simultaneously to the incident longitudinal and transverse electric fields 
(fundamentally represented by the scalar and vector potentials, respectively, under the Coulomb gauge),
whereas the latter system is exposed to only the transverse field (the vector potential under the Coulomb gauge).}
\label{fig:NFOvsOO}
\end{center}
\end{figure}
Suppose a small-scale material is placed in the vicinity of a nanostructure, which functions as a light source (\fig{fig:NFOvsOO}).
In such a system, under the NF incidence condition, the target material is exposed to  
the longitudinal and transverse electric fields simultaneously, whereas in 
a system under the far-field incidence condition, the target material is 
exposed only to the transverse field, 
which survives far from the light source.
Therefore, the coexistence of longitudinal and transverse electric fields distinguishes 
such a system under the NF incidence condition from that under the far-field incidence condition.  

Here, the longitudinal electric field originates from the charge 
density on the nanostructure, obeys Coulomb's law, and  
has a non-radiative nature to localize around the nanostructure.
On the other hand, the transverse electric field originates from 
the transverse current density on the nanostructure, obeys 
Ampere-Maxwell law and Faraday's law, and 
has a radiative nature allowing it to propagate far from the light source, accompanied by the magnetic field.
(The longitudinal current density is determined via the charge conservation law, once the charge density is known, and is not an independent source.)
Therefore, the two incidences coexisting in an NF optical system 
 have distinct properties. 

Furthermore, owing to {\it the non-relativistic nature} of the system,
the SP and VP appear in a different manner in the Hamiltonian, 
(for example, \eq(H) in \S\ref{sec:susHop},) which governs the electron response. 
Considering that the SP and VP under the Coulomb gauge 
represent the longitudinal and transverse electric fields, respectively, 
one may confirm that 
the two types of incidences in NFO cause different responses
; see \S\ref{sec:oneElectron} for an explicite demonstration.

Therefore, it is reasonable to treat SP and VP equally as the cause of response 
in  the NF optical system.
Up to now, there has been no such theoretical framework
for equally treating the SP and VP.
The reason for this lies in the the many-electron problem inevitably related to NFO
via the SP (the longitudinal electric field), as is mentioned in the next subsection.
%
\subsection{A many-electron problem inevitably related with NFO}
\label{sec:MEP}
%
The relationship between NFO and many-electron problem
has not been well recognized,  
although the problem of how to consider  the Coulomb interaction 
in response function has remained 
for a long time\cite{Toyozawa}.
In the usual Hamiltonian for a many-electron system under the Coulomb gauge,
the SP is rewritten as the interaction 
between the electron charge density operators, 
and only the VP is considered as the cause of the response.
This unequal treatment of the SP and VP
is needed to consider the quantum many-electron effect (the so-called exchange-correlation effect)
to construct the ground and excited states 
as the proper bound states in a many-electron system.
This usual procedure to treat the non-relativistic many-electron system is compatible with 
ordinary optics, 
where the electron system of interest is far from 
the light source, and the SP incidence is negligible.
By contrast, in an NF optical system, 
the usual approach results in a difficulty of understanding the response to 
the SP incidence,
because both the SP incidence (radiated by the nanostructure) and
the inherent SP (originating from the particle charge) are 
built into the two-body Coulomb interaction, 
and the two contributions are indistinguishable.
To make matters worse, the Coulomb interaction in itself is so difficult to treat 
that it is often ignored, without considering it includes the effect of the 
SP incidence.

For the NF optical system, there are two existing approaches based on certain single susceptibilities 
(the nonlocal response functions).
Cho formulated a single susceptibility that relates the transverse VP (as the cause) 
to the current density	 (as the result), and applied it to various optical systems\cite{Cho00}.
Additionally, a modification that considers the longitudinal electric field incidence 
in NF optical systems has been proposed in Chap. 5 of Ref.\cite{Cho02}. 
Keller formulated the linear single susceptibility, which relates 
the transverse electric field and the incident part of the longitudinal electric field
(as the cause) to the current density (as the result) \cite{Keller01}.

In the above two existing formulations, the SP under the Coulomb gauge 
(or the longitudinal electric field), 
except the linear-dependence of the incidence, is rewritten as the two-body Coulomb interaction
in the usual manner.
Therefore, the response to the SP, in principle, can be 
rigorously considered via the Coulomb interaction if the many-electron problem is properly solved, 
whereas the response to the VP incidence under the Coulomb gauge 
(or the transverse electric field incidence) is treated in the perturbative manner. 
In this type of approach, it is essential to solve the many-electron problem, in particular, for the nonlinear process related with the SP.  
Even if the Coulomb interaction is properly considered, 
the unequal treatment may make it difficult to regulate the perturbation order of the responses 
and to understand the role of the SP incidence.

As a result, the response theory in NFO is inevitably relates to the many-electron problem, 
which causes difficulty.
\subsection{The purpose of this paper}
\label{sec:purpose}
\S\ref{sec:singleSus}-\S\ref{sec:MEP} lead to the logical fallacy to use
the ordinary two susceptibilities with the electric and magnetic fields, 
and the preference to use the single susceptibility equally associated with the SP and VP,
considering properly the many-electron effect in NF optical systems,  
although the ordinary two susceptibilities have been widely used 
both in ordinary optics and NFO. 
To best understand the fundamental physics in NFO, 
it is essential to develop an adequate response theory.
For this purpose, the present paper defines and characterizes a single susceptibility 
equally associated with the SP and VP
based on the action integral from scratch.

 The contents of this paper are as follows:
\S\ref{sec:definition} defines the linear and nonlinear single susceptibilities
equally associated with the SP and VP, 
starting from the action integral.
 \S\ref{sec:ccl_gi} shows that the present susceptibility respects 
both charge conservation and gauge invariance in a general manner. 
\S\ref{sec:susHop} derives the linear and nonlinear single susceptibilities
in the form of the Heisenberg operators. 
\S\ref{sec:eigenstate} shows that 
the many-electron effect in the present response theory may be supported by
the density functional theory to prepare the non-perturbed ground state as well as a complete set
of many-electron states.
\S\ref{sec:oneElectron} applies the present linear response theory to a simplified 
many-electron system,
and show that the electric field with the associated permittivity is not suitable 
to describe the response of a limited NF optical system with a non-metallic material, 
so that the SP and VP with the single susceptibility is essential.
\S\ref{sec:summary} provides a summary of this work.
Two appendices are included: 
\S\ref{sec:derivativeZero} and  \S\ref{sec:methods} provides calculation details of 
\S\ref{sec:definition} and \S\ref{sec:oneElectron}, respectively.
%
%
\section{Definition of New Single Susceptibility}
\label{sec:definition}
%
Based on the Lagrangian formulation of non-relativistic quantum electrodynamics, 
we define the single susceptibility, which relates the SP and VP (as the cause) to
the induced charge and current densities (as the result). Furthermore, it is
shown that this susceptibility guarantees that charge conservation and gauge invariance hold; 
see the next section.
The action integral for non-relativistic quantum electrodynamics is:
\begin{align}
\label{action}
\c{I}[\h{\psi}_{\alpha}^{\dagger},\h{\psi}_{\alpha},\phi,\v{A}] 
&\equiv
\c{I}_{\text{mat}}[\h{\psi}_{\alpha}^{\dagger},\h{\psi}_{\alpha},\phi,\v{A}]
+ \c{I}_{\text{EM}}[\phi,\v{A}],
\\
\nonumber
\c{I}_{\mbox{mat}}[\h{\psi}_{\alpha}^{\dagger},\h{\psi}_{\alpha},\phi,\v{A}] 
&\equiv
 \frac{1}{c}\int \dn{4}{x} 
 \left\{
 \h{\psi}_{\alpha}^{\dagger}(x)( i\hbar \del{t} -q\phi(x) ) \h{\psi}_{\alpha}(x)
 \right.
\\
\nonumber
&\quad
- \frac{1}{2m} 
\left(   \frac{\hbar}{-i} \del{i} -qA_{i}(x)  \right) \h{\psi}_{\alpha}^{\dagger}(x) \cdot
\left(   \frac{\hbar}{i} \del{i} -qA_{i}(x)  \right) \h{\psi}_{\alpha}(x)
\\
\label{action_mat}
&\quad
\left.
 -\phi(x) \rho^{\text{(EXT)}}(x) + A_{i}(x) j_{i}^{\text{(EXT)}}(x)
 - \h{\psi}_{\alpha}^{\dagger}(x) v^{\text{(AUX)}}(x)  \h{\psi}_{\alpha}(x)
\right\}
\\
\nonumber
\c{I}_{\text{EM}}[\phi,\v{A}]
&\equiv 
\frac{1}{c}\int \dn{4}{x} \left\{
\frac{\evac}{2}
\left(  \del{t}A_{i}(x) + \del{i} \phi(x)\right) \left(  \del{t}A_{i}(x) + \del{i} \phi(x)\right)
\right. 
\\ 
\label{action_EM}
&\quad
\left.
-\frac{\evac c^{2}}{2} 
\epsilon_{ijk} \del{j} A_{k}(x)  \epsilon_{ilm} \del{l} A_{m}(x)
\right\},
\end{align}
where 
$m$ and $q(=-e)$ are the electron mass and charge, $c$ is the speed of light, 
$\phi,\v{A}$ are the SP and VP, which are assumed to be classical field in the semiclassical treatment, 
$\h{\psi}_{\alpha}^{\dagger},\h{\psi}_{\alpha}$ are the electron field operators 
with the spin state $\alpha$ (one of the two spin states; so called "up" and "down" states), and 
$\rho^{\text{(EXT)}},\v{j}^{\text{(EXT)}}$ are the nuclear charge and the current densities,
respectively, which possibly generate inherent EM field.
A static auxiliary potential $v^{\text{(AUX)}}(x) $ is null for now, 
but is introduced here for the discussion in \S\ref{sec:eigenstate} concerning 
the density functional theory
to consider the quantum many-electron effect (the exchange-correlation effect),
$\epsilon_{ijk} $ is an antisymmetric tensor, and the Einstein rule is 
used for indices of vector and Grassmann fields, that is,
summation should be executed over repeated indices. 
At this first stage of investigation, 
the interaction between spin polarization and the EM field is ignored.
The soundness of the above action integral is confirmed by its Euler equations, which will soon be
derived.

The electron field operators are considered as quantized Grassmann fields.
The Grassmann field satisfies 
$[ \h{\psi}_{\alpha}(\v{r},t), \h{\psi}_{\beta}^{\dagger}(\v{r}^{\prime},t^{\prime}) ]_{+} = 0$
\cite{Grassmann},
and corresponds to the "classical" field of the electron.
These operators become the creation and annihilation operators of the electron in quantum theory
(the quantized Grassmann fields),  
introducing the anti-commutation relationship:
$[ \h{\psi}_{\alpha}(\v{r},t), \h{\psi}_{\beta}^{\dagger}(\v{r}^{\prime},t) ]_{+} 
= \delta^{3}(\v{r}-\v{r}^{\prime})\delta_{\alpha \beta}$.

The action integral is composed of two parts:
one is the action integral of the matter (including the interaction between the matter and EM field), 
$\c{I}_{\mbox{mat}}[\h{\psi}_{\alpha}^{\dagger},\h{\psi}_{\alpha},\phi,\v{A}]$,
and the other is the action integral of the EM field, 
$\c{I}_{\text{EM}}[\phi,\v{A}]$.
Applying the extremal (optimizing) conditions with respect to $\h{\psi}_{\alpha}(x)\,,\h{\psi}_{\alpha}^{\dagger}(x)$
leads to Heisenberg's equation, and optimizing
with respect to $\phi(x),\, \v{A}(x)$
leads to Maxwell's wave equations:
\beqna
\nonumber
0&=&c\: 
\delta\h{\psi}_{\alpha}^{\dagger}(x)  {\Large\backslash} \delta\c{I} 
=
c\:
\delta\h{\psi}_{\alpha}^{\dagger}(x)  {\Large\backslash} \delta\c{I}_{\mbox{mat}} 
\\
\label{psi_opt}
&=&
\left( i\hbar \del{t}   -q\phi(x)    - \frac{1}{2m} 
\left(   \frac{\hbar}{i} \del{i} -qA_{i}(x)  \right) \cdot
\left(   \frac{\hbar}{i} \del{i} -qA_{i}(x)  \right) -v^{\text{(AUX)}}(x) \right) \h{\psi}_{\alpha}(x),
\\
\nonumber
0&=&c\: 
\delta\c{I}{\Large /}\delta\h{\psi}_{\alpha}(x)
=
c\:
\delta\c{I}_{\mbox{mat}}{\Large /}\delta\h{\psi}_{\alpha}(x)
\\
\label{psiDagger_opt}
&=&
\left( -i\hbar \del{t}   -q\phi(x)    - \frac{1}{2m} 
\left(   \frac{\hbar}{-i} \del{i} -qA_{i}(x)  \right) \cdot
\left(   \frac{\hbar}{-i} \del{i} -qA_{i}(x)  \right) -v^{\text{(AUX)}}(x) \right) \h{\psi}_{\alpha}^{\dagger}(x),
\\
\label{A_opt}
0&=&c\: \frac{\delta \c{I}}{\delta A_{i}(x)}=\evac c^{2} 
\left( -\epsilon_{ijk}\del{j} \epsilon_{klm} \del{l}A_{m}(x) -\frac{1}{c^{2}}\del{t}^{2} A_{i}(x) 
-\frac{1}{c^{2}}\del{t} \del{i} \phi(x) 
 +   \frac{1}{\evac c^{2}} ( \h{j}_{i} (x) +j_{i}^{\text{(EXT)}}(x) ) \right),
\\
\label{phi_opt}
0&=&c\: \frac{\delta \c{I}}{\delta \phi(x)}=\evac 
\left( -\del{i}\del{i} \phi(x) -\del{t} \del{i} A_{i}(x) -  \frac{1}{\evac} ( \h{\rho}(x) + \rho^{\text{(EXT)}}(x) )\right) .
\eeqna
In \eqs(psi_opt)-(psiDagger_opt), the left- and right-hand functional derivatives with respect to the Grassmann field are executed, respectively.
In \eqs(A_opt)-(phi_opt), the following
definitions are introduced for the electron charge and current densities,
respectively:
\beqna
\label{charge00}
\h{\rho}(x) &\equiv& -c \frac{\delta}{\delta \phi(x)} \c{I}_{\mbox{mat}}
= q \h{\psi}_{\alpha}^{\dagger}(x) \h{\psi}_{\alpha}(x),
\\
\label{current00}
\h{j}_{i}(x)&\equiv& 
+c \frac{\delta}{\delta A_{i}(x)} \c{I}_{\mbox{mat}}
=
\frac{q}{2m} 
 \h{\psi}_{\alpha}^{\dagger}(x) \left(   \frac{\hbar}{i} \del{i} -qA_{i}(x)  \right) \h{\psi}_{\alpha}(x) + \mbox{h.c.}\,.
\eeqna
The charge conservation law below holds, and is checked through explicit calculation:
\beq
\label{ccl00}
\partial_{t}\h{\rho}(x) + \partial_{i}\h{j}_{i}(x) =0.
\eeq 
In the four-element representation, 
\eqs(A_opt)-(phi_opt) become:
\begin{align}
\label{Meq_4}
& \left(\delta^{\mu}_{\;\;\nu} \Box - \partial^{\mu}\del{\nu} \right) A^{\nu}(x)
= \frac{1}{\evac c}( \h{j}^{\mu}(x) +j^{\text{(EXT)}\mu}(x)),
\\
\text{where}\quad
\nonumber
&\h{j}^{\mu} = (c\h{\rho}, \hv{j}),\:\h{j}_{\mu} = (c\h{\rho}, -\hv{j}),\:
\\
\nonumber
&A^{\mu} = (\phi, c\v{A}),\:A_{\mu} = (\phi, -c\v{A}),\:
\\
\nonumber
&\partial^{\mu}=(1/c\: \del{t}, -\grad),\: \del{\mu} = (1/c\: \del{t}, \grad),\:
\\
\label{4element}
&\Box =  \partial^{\mu}\del{\mu} = 1/c^{2}\:\del{t}^{2} - \Delta \; \text{\;, etc.} 
\end{align}
Although Lorentz invariance is not maintained in the non-relativistic theory,
we use the four-element notation to simply represent charge conservation and gauge invariance.
For example, \eqss(charge00)-(ccl00) become:
\begin{align}
\label{current_4}
& \h{j}^{\mu} (x) =  -c^{2} \frac{\delta}{\delta A_{\mu}(x)} \c{I}_{\mbox{mat}},
\\
\label{ccl_4}
& \partial_{\mu}\,\h{j}^{\mu}(x) =0.
\end{align}

The action integral, \eq(action) is invariant under the following gauge transformation:
\beqna
\nonumber
A^{\mu}(x) &\to& A^{\mu}(x) - c\,\partial^{\mu} \eta(x),
\\
\label{gt}
\h{\psi}_{\alpha}(x) &\to& e^{\frac{i}{\hbar} q \eta(x)}\h{\psi}_{\alpha}(x),\quad
\h{\psi}_{\alpha}^{\dagger}(x) \to \h{\psi}_{\alpha}^{\dagger}(x) e^{\frac{-i}{\hbar} q \eta(x)}\,,
\eeqna
where $\eta(x)$ is the gauge function.
From the point of view of Noether's theorem\cite{EDM2},
the gauge invariance of the action integral is the cause of the charge conservation law, 
\eq(ccl00) or \eq(ccl_4).

Let us separate the EM field into two parts:
\beq
\label{EMpotSeparated}
A^{\mu}(x) = A^{(0)\mu}(x) + \Delta A^{\mu}(x),
\eeq
where $A^{(0)\mu}$ is the static, non-perturbative EM potential satisfying
\eqs(A_opt)-(phi_opt), and $\Delta A^{\mu}(x)$ is the perturbative EM potential.
Under this variation of the EM field, let us re-optimize the action integral of the matter, 
$\c{I}_{\mbox{mat}}[\h{\psi}_{\alpha}^{\dagger},\h{\psi}_{\alpha},A^{\mu}]$.
That is, we re-optimize the electron field operator satisfying \eqs(psi_opt)-(psiDagger_opt)
under $A^{(0)\mu} + \Delta A^{\mu}(x)$.
In the above procedure, the variation of the action integral of the matter is expressed by
the total functional derivative with respect to $A^{\mu}(x)$:
\beqna
\nonumber
&&
\left.
\frac{\delta}{\delta A_{\mu}(x)}\c{I}_{\mbox{mat}} 
[\h{\psi}_{\alpha}^{\dagger}[A^{\nu}]\,,\h{\psi}_{\alpha}[A^{\nu}]\,,A^{\nu}]
\right|_{A^{\nu}=A^{(0)\nu}}
\\
\nonumber
&=&
\left[ \left. \frac{\delta}{\delta A_{\mu}(x)} \right|_{\mbox{{\small explicit}}} \c{I}_{\mbox{mat}} 
+ \int \dn{4}{x}^{\prime}\frac{\delta \h{\psi}_{\alpha}^{\dagger}(x^{\prime})}{\delta A_{\mu}(x)}
\delta \h{\psi}_{\alpha}^{\dagger}(x^{\prime}) {\Large \backslash} \delta\c{I}_{\mbox{mat}}
\right.
\\
\nonumber
&&
\left.
\hspace{0.25\textwidth}
+ \int \dn{4}{x}^{\prime} \delta\c{I}_{\mbox{mat}} {\Large /} \delta \h{\psi}_{\alpha}(x^{\prime}) 
\frac{\delta \h{\psi}_{\alpha}(x^{\prime})}{\delta A_{\mu}(x)}
\right]_{A^{\nu}=A^{(0)\nu}}
\\
\label{firstFunctionalDerivative}
&=&\frac{-1}{c^{2}}\h{j}^{\mu}(x;[A^{(0)\nu}])\,,
\eeqna
where the first term in the second expression is the variation explicitly caused by 
the perturbative EM field, and the second and third terms are the implicit variations,
created through re-optimization of the field operator to satisfy 
 \eqs(psi_opt)-(psiDagger_opt) under the existence of the perturbative EM field.
The last expression is derived using \eq(current_4), \eqs(psi_opt)-(psiDagger_opt).
The above equation reveals that the first order total functional derivative of the action integral of the matter 
is simply the current density in the non-perturbed system.  Furthermore, the second order total functional derivative
is calculated as follows:
\beqna
\nonumber
&&
\left.
\frac{\delta}{\delta A^{\mu_{1}}(x_{1})} 
\frac{\delta}{\delta A_{\mu}(x)} \c{I}_{\mbox{mat}} 
[\h{\psi}_{\alpha}^{\dagger}[A^{\nu}]\,,\h{\psi}_{\alpha}[A^{\nu}]\,,A^{\nu}]
\right|_{A^{\nu}=A^{(0)\nu}}
\\
\nonumber
&=&
\left[
\frac{\delta}{\delta A^{\mu_{1}}(x_{1})}
\left( \left. \frac{\delta}{\delta A_{\mu}(x)} \right|_{\mbox{{\small explicit}}} \c{I}_{\mbox{mat}} \right)
+
 \int \dn{4}{x}^{\prime} \frac{\delta}{\delta A^{\mu_{1}}(x_{1})}
\left( \frac{\delta \h{\psi}_{\alpha}^{\dagger}(x^{\prime})}{\delta A_{\mu}(x)} 
\delta \h{\psi}_{\alpha}^{\dagger}(x^{\prime}) {\Large \backslash} \delta\c{I}_{\mbox{mat}} \right)
\right.
\\
\nonumber
&&
\left.
\hspace{0.395\textwidth}
+ \int \dn{4}{x}^{\prime} \frac{\delta}{\delta A^{\mu_{1}}(x_{1})}
\left( \delta\c{I}_{\mbox{mat}} {\Large /} \delta \h{\psi}_{\alpha}(x^{\prime}) 
\frac{\delta \h{\psi}_{\alpha}(x^{\prime})}{\delta A_{\mu}(x)} \right)
\right]_{A^{\nu}=A^{(0)\nu}}
\\
\label{secondFunctionalDerivative}
&=&
\frac{-1}{c^{2}}
\left. \frac{\delta \h{j}^{\mu}(x;[A^{\nu}])}{\delta A^{\mu_{1}}(x_{1})} 
\right|_{A^{\nu}=A^{(0)\nu}} ,
\eeqna
where the second and third terms in the second expression are null.
Actually, the integrand of the second term is:
\[
\left[
\left( \frac{\delta}{\delta A^{\mu_{1}}(x_{1})}
\frac{\delta \h{\psi}_{\alpha}^{\dagger}(x^{\prime})}{\delta A_{\mu}(x)} \right)  
\delta \h{\psi}_{\alpha}^{\dagger}(x^{\prime}) {\Large \backslash} \delta\c{I}_{\mbox{mat}} 
+
\frac{\delta \h{\psi}_{\alpha}^{\dagger}(x^{\prime})}{\delta A_{\mu}(x)} 
\left( \frac{\delta}{\delta A^{\mu_{1}}(x_{1})}
\delta \h{\psi}_{\alpha}^{\dagger}(x^{\prime}) {\Large \backslash} \delta\c{I}_{\mbox{mat}}\right)
\right]_{A^{\nu}=A^{(0)\nu}}\,,
\]
The first term in this equation is null because of \eq(psi_opt),
and the second term is also null because of \eq(nFunctionalDerivativeA02) 
in Appendix \ref{sec:derivativeZero}. 
In the same manner as for higher order total functional derivatives of the action integral of the matter, 
the following extension of \eq(secondFunctionalDerivative)
holds, owing to \eqs(nFunctionalDerivativeA01)-(nFunctionalDerivativeA02) in
Appendix \ref{sec:derivativeZero},
\beq
\label{nFunctionalDerivative}
\left. \frac{\delta^{n+1}\c{I}_{\mbox{mat}} 
[\h{\psi}_{\alpha}^{\dagger}[A^{\nu}]\,,\h{\psi}_{\alpha}[A^{\nu}]\,,A^{\nu}]
}{
\delta A^{\mu_{n}}(x_{n})\cdots 
\delta A^{\mu_{1}}(x_{1}) \delta A_{\mu}(x)} 
\right|_{A^{\nu}=A^{(0)\nu}} 
=
\frac{-1}{c^{2}}
\left. \frac{\delta^{n} \h{j}^{\mu}(x;[A^{\nu}])}{\delta A^{\mu_{n}}(x_{n}) \cdots \delta A^{\mu_{1}}(x_{1})} 
\right|_{A^{\nu}=A^{(0)\nu}}\,.
\eeq

To define the single susceptibility, 
suppose the system under the non-perturbative EM field $A^{(0)\mu}(x)$
is exposed to the perturbative EM field $\Delta A^{\mu}(x)$.
The non-perturbative EM field $A^{(0)\mu}$ is a solution of the coupled equations \eqss(psi_opt)-(phi_opt), namely,
Heisenberg's equation and Maxwell's wave equations, and is assumed to be a static solution 
existing in the ground state.
On the other hand, the total EM field  $A^{(0)\mu}+ \Delta A^{\mu}$
is not necessarily a solution of Maxwell's wave equations, \eqs(A_opt)-(phi_opt), 
that is, $\Delta A^{\mu}$ is introduced as a virtual variation.
The induced current density is the variation from the current density in
the non-perturbative system:
\beqna
\nonumber
&&\h{j}^{\mu} (x; [A^{(0)\nu} + \Delta A^{\nu}]) -  \h{j}^{\mu} (x; [A^{(0)\nu} ])
\\
\nonumber
&=&\quad
\int \dn{4}{x_{1}} \left. \frac{\delta \h{j}^{\mu}(x;[A^{\nu} ])) }{ \delta A^{\mu_{1}}(x_{1})} 
\right|_{A^{\nu}=A^{(0)\nu}}  
\hspace{-1.0cm}
\Delta A^{\mu_{1}}(x_{1}) 
\\
\nonumber
&+& \frac{1}{2!}
\int \dn{4}{x_{1}} \int \dn{4}{x_{2}} 
\left.
\frac{\delta^{2} \h{j}^{\mu}(x;[A^{\nu} ]) }{ \delta A^{\mu_{1}}(x_{1}) \delta A^{\mu_{2}}(x_{2})} 
\right|_{A^{\nu}=A^{(0)\nu}} 
\hspace{-1.0cm}
\Delta A^{\mu_{1}}(x_{1}) \Delta A^{\mu_{2}}(x_{2}) 
\\
\nonumber
&+& \frac{1}{3!}
\int \dn{4}{x_{1}} \int \dn{4}{x_{2}}  \int \dn{4}{x_{3}} 
\left.
\frac{\delta^{3} \h{j}^{\mu}(x;[A^{\nu} ]) }{ \delta A^{\mu_{1}}(x_{1}) \delta A^{\mu_{2}}(x_{2}) 
\delta A^{\mu_{3}}(x_{3})} \right|_{A^{\nu}=A^{(0)\nu}} 
\hspace{-1.0cm}
\Delta A^{\mu_{1}}(x_{1}) \Delta A^{\mu_{2}}(x_{2}) \Delta A^{\mu_{3}}(x_{3})
\\
\label{jind}
&+& \cdots\, .
\eeqna
From \eqs(nFunctionalDerivative)-(jind), 
the linear and nonlinear single susceptibility operators are defined as:
\beqna
\nonumber
\h{\chi}^{\mu}_{\;\;\mu_{1} }(x,x_{1}) &\equiv&
\left. 
\frac{\delta \h{j}^{\mu}(x;[A^{\nu}])}{\delta A^{\mu_{1}}(x_{1})} 
\right|_{A^{\nu}=A^{(0)\nu}}
\\
\label{resp01}
&=&
\left. -c^{2}
\frac{\delta^{2} \c{I}_{\mbox{mat}}}{ \delta A_{\mu}(x) \delta A^{\mu_{1}}(x_{1})} 
\right|_{A^{\nu}=A^{(0)\nu}}\;,
\\
\nonumber
\h{\chi}^{\mu}_{\;\;\mu_{1}\,\mu_{2}}(x,x_{1},x_{2}) &\equiv&
\frac{1}{2!}
\left. 
\frac{\delta^{2} \h{j}^{\mu}(x;[A^{\nu}])}
{\delta A^{\mu_{1}}(x_{1}) \delta A^{\mu_{2}}(x_{2}) }
\right|_{A^{\nu}=A^{(0)\nu}}\;,
\\
\label{resp02}
&=&
\frac{-c^{2}}{2!}
\left. 
\frac{\delta^{3}  \c{I}_{\mbox{mat}}}
{ \delta A_{\mu}(x) \delta A^{\mu_{1}}(x_{1}) \delta A^{\mu_{2}}(x_{2}) }
\right|_{A^{\nu}=A^{(0)\nu}}
\\
\nonumber
\h{\chi}^{\mu}_{\;\;\mu_{1}\,\cdots\mu_{n}}(x,x_{1},\cdots,x_{n}) &\equiv&
\frac{1}{n!}
\left. 
\frac{\delta^{n} \h{j}^{\mu}(x;[A^{\nu}])}
{\delta A^{\mu_{1}}(x_{1}) \cdots \delta A^{\mu_{n}}(x_{n})}
\right|_{A^{\nu}=A^{(0)\nu}}
\\
\label{resp0n}
&=&
\frac{-c^{2}}{n!}
\left. 
\frac{\delta^{n+1} \c{I}_{\mbox{mat}}}
{ \delta A_{\mu}(x) \delta A^{\mu_{1}}(x_{1}) \cdots \delta A^{\mu_{n}}(x_{n})}\right|_{A^{\nu}=A^{(0)\nu}}
\;,
\eeqna
%
%
%

The susceptibility is defined using a small amount of variation, $\Delta A^{\mu}$.
That is, the EM field does not in general satisfy its Euler equation, \eq(Meq_4), while
the electron field operators satisfy  \eqs(psi_opt)-(psiDagger_opt).
To evaluate the real EM field, $\Delta A^{\mu}$ must be determined
and a further procedure is required to solve the coupled equations, with 
the constitutive equations in terms of the susceptibility and 
Maxwell's wave equations \eqs(A_opt)-(phi_opt). This procedure is provided in a self-consistent manner,
as performed by K.Cho\cite{Cho00} using his single susceptibility.
%
\section{Charge Conservation Law and Gauge Invariance of the Single Susceptibility}
\label{sec:ccl_gi}
%
In the last expressions in \eqss(resp01)-(resp0n) 
the coordinates $x_{1},x_{2},\cdots$ for the cause (the perturbative EM field)
and the coordinates $x$ for the result (the induced current density) are
symmetric. 
Charge conservation for the induced charge density 
holds to each order of the perturbation
because of \eq(ccl00) or \eq(ccl_4) and \eqss(jind)-(resp0n);
this is described 
by the derivative of  the coordinate for the result, $x$:
\beq
\label{ccl}
\partial_{\mu} \h{\chi}^{\mu}_{\;\;\mu_{1} \cdots }(x,x_{1}, \cdots) =0\,.
\eeq
The symmetry of the coordinates between the result and the cause
leads to the following equation 
concerning the derivative of any coordinate for the cause, e.g., $x_{1}$ :
\beq
\label{gi}
\partial^{\mu_{1}} \h{\chi}^{\mu}_{\;\;\mu_{1} \cdots }(x,x_{1}, \cdots) =0 .
\eeq
\Eq(gi) means that the susceptibility guarantees that gauge invariance is respected. That is,
the resultant charge and current densities are independent
of the chosen gauge.
To confirm this fact, consider the convolution integral 
of the single susceptibility with the perturbative EM field,
in a certain gauge, e.g.,
\beq
\label{convolution01}
\int \dn{4}{x_{1}}\;  \h{\chi}^{\mu}_{\;\;\mu_{1} \cdots }(x,x_{1}, \cdots)
\Delta A^{\mu_{1}}(x_{1}) .
\eeq
A gauge transformation of  $\Delta A$ to $\Delta A^{\prime}$ in another gauge is expressed as :
\beq
 \Delta A^{\mu_{1}}(x_{1}) = \Delta A^{\prime \mu_{1}}(x_{1}) - c\, \partial^{\mu_{1}} \eta(x_{1}),
\eeq
where $\eta$ is the gauge function.
\Eq(convolution01) leads to:
\beqna
\nonumber
&&
\int \dn{4}{x_{1}}\;  \h{\chi}^{\mu}_{\;\;\mu_{1} \cdots }(x,x_{1}, \cdots)
\Delta A^{\mu_{1}}(x_{1}) 
\\
\nonumber
&=&
\int \dn{4}{x_{1}}\;  \h{\chi}^{\mu}_{\;\;\mu_{1} \cdots }(x,x_{1}, \cdots)
\Delta A^{\prime \mu_{1}}(x_{1}) + c
\int \dn{4}{x_{1}}\;  \partial^{\mu_{1}} \h{\chi}^{\mu}_{\;\;\mu_{1} \cdots }(x,x_{1}, \cdots)
\eta(x_{1})  
\\
\label{convolution02}
&=&\int \dn{4}{x_{1}}\;  \h{\chi}^{\mu}_{\;\;\mu_{1} \cdots }(x,x_{1}, \cdots)
\Delta A^{\prime \mu_{1}}(x_{1}) .
\eeqna
The contribution of the gauge function vanishes in the convolution integral. 
Thus, the gauge of the perturbative EM field may be freely selected.
This means that the susceptibility is independent of the chosen gauge and,
in practice, one may select a gauge that is most convenient for calculation.
%
\section{Single Susceptibility in the form of Heisenberg Operator}
\label{sec:susHop}
%
In this section, the linear and nonlinear single susceptibilities
in the form of Heisenberg operators are derived 
using an expansion of the retarded product in Hamiltonian formulation\cite{Nishijima}.
The Heisenberg operator of four-element current density, i.e., 
$\h{j}^{\mu}(x) = (c \h{\rho}(x),\,\hv{j}(x) )$ is:
\beqna
\label{j4_H}
\h{j}^{\mu}(x)&=&
\left\{
\begin{array}{lll}
cq  \h{\psi}_{\alpha}^{\dagger}(x) \h{\psi}_{\alpha}(x) & \text{ for } & \mu=0,
\\
\displaystyle
\h{\psi}_{\alpha}^{\dagger}(x)  \frac{q}{2m} \left( \frac{\hbar}{i} (-\partial^{\mu}) -\frac{q}{c}A^{\mu}(x) \right) \h{\psi}_{\alpha}(x) + \mbox{h.c.} &\text{ for }  &\mu = 1,2,3\; .
\end{array}
\right.
\eeqna
In \eq(action_mat), if the factor $i\hbar \h{\psi}_{\alpha}^{\dagger}(x)$ of the first term is regarded as
the canonical momentum of $\h{\psi}_{\alpha}(x)$,
then the Hamiltonian density may be determined
as the Legendre transformation from the Lagrangian density, that is:
\begin{align}
\label{H}
\h{H} \quad
&\equiv
\int \dn{3}{x}\;
\frac{1}{2m}
\left(\frac{\hbar}{-i} \del{i} -qA_{i}(x) \right)\h{\psi}_{\alpha}^{\dagger}(x)\:
\left(\frac{\hbar}{i} \del{i} -qA_{i}(x) \right)\h{\psi}_{\alpha}(x)
+ q\phi(x) \:\h{\psi}_{\alpha}^{\dagger}(x)\h{\psi}_{\alpha}(x).
\end{align}
This Hamiltonian  governs the motion of electron field operators.
Assuming that the non-perturbative EM field $\phi^{(0)}, \v{A}^{(0)}$ is
the static EM field existing in the ground state of a many-electron system,
the Hamiltonian, $\h{H}$ may be separated into a non-perturbative part, $\h{H}^{(0)}$ 
and a perturbative part, $\h{V}$ as follows: 
\begin{align}
\nonumber
 \h{H}^{(0)}
&\equiv
\int \dn{3}{x}\;
\frac{1}{2m}
\left(\frac{\hbar}{-i} \del{i} -qA^{(0)}_{i}(x) \right)\h{\psi}_{\alpha}^{\dagger}(x)\cdot
\left(\frac{\hbar}{i} \del{i} -qA^{(0)}_{i}(x) \right)\h{\psi}_{\alpha}(x)
+ q\phi^{(0)}(x) \:\h{\psi}_{\alpha}^{\dagger}(x)\h{\psi}_{\alpha}(x)
\\
\label{H0}
&\quad
+ v^{\text{(AUX)}}(x) \:\h{\psi}_{\alpha}^{\dagger}(x)\h{\psi}_{\alpha}(x),
\\
\nonumber
\h{V}(t) 
&\equiv  \h{H}- \h{H}^{(0)} = \int \dn{3}{x} \h{v}(x),
\\
\nonumber
&=
\int \dn{3}{x}\left\{
\Delta\phi(x)\:q \h{\psi}_{\alpha}^{\dagger}(x)\h{\psi}_{\alpha}(x)
- \Delta A_{i}(x)\,\frac{q}{2m} \left(\h{\psi}_{\alpha}^{\dagger}(x)\:
\left(\frac{\hbar}{i} \del{i} -qA^{(0)}_{i}(x) \right)\h{\psi}_{\alpha}(x) + \mbox{h.c.} \right) 
\right.
\\
\nonumber
&\hspace{0.1\textwidth}
\left.
+\frac{q}{2m}  \Delta A_{i}(x)  \Delta A_{i}(x) 
\:q \h{\psi}_{\alpha}^{\dagger}(x)\h{\psi}_{\alpha}(x)
\right\}
\\
\label{Vdensity}
&= 
\int \dn{3}{x}
\left\{
\frac{1}{c}\Delta A^{\mu}(x) \left.\h{j}_{\mu}(x)\right|_{A=A^{(0)}}  
-\frac{q}{2mc^{3}}\,
\t{\delta}_{\mu}^{\;\mu^{\prime}}
\Delta A^{\mu}(x) \Delta A_{\mu^{\prime}}(x) \h{j}_{0}(x) 
\right\},
\\
\label{nonrel_delta}
&\text{\quad where } 
\t{\delta}_{\mu}^{\;\mu^{\prime}}  = 
\left\{
 \begin{array}{ll}
 1 & \text{ for } \mu=\mu^{\prime}=1,2,3\:,\\
 0 & \text{ otherwise}\,.
\end{array}
\right. 
\end{align}
The auxiliary potential, $v^{\text{(AUX)}}(x)$ effectively represents for
the quantum many-electron effect (the exchange-correlation effect); 
this fact will be explained in the next section. 
The factor $\displaystyle \left.\h{j}_{\mu}(x)\right|_{A=A^{(0)}}$ in \eq(Vdensity)
is the current density \eq(j4_H), with the explicitly-appeared VP being replaced by that in the non-perturbed system.
The tensor \eq(nonrel_delta) represents the non-relativistic effect.
Actually, this tensor is the analogue of the four-element Kronecker delta, 
but brings inequality of the temporal and spatial coordinates.

Here, the field operators in the interaction picture (the asymptotic field operators) 
$\h{\psi}_{\alpha}^{(in)\dagger}, \h{\psi}_{\alpha}^{(in)}$ are governed by 
the non-perturbative Hamiltonian $\h{H}^{(0)}$
and coincide with the field operators in the Heisenberg picture,
$\h{\psi}_{\alpha}^{\dagger}, \h{\psi}_{\alpha}$
at the infinite past time, $t \to -\infty$,
assuming the adiabatic switch-on.
The unitary operator $\h{U}(t,-\infty)$ is the time-evolution operator of the states in the interaction picture,
and relates the operators between the Heisenberg and interaction pictures as follows:
\beqna
\label{def_psi_I}
\h{\psi}_{\alpha}(x)&=&\h{U}^{-1}(t,-\infty)\h{\psi}_{\alpha}^{(in)}(x)\h{U}(t,-\infty),
\\
\nonumber
\h{\psi}_{\alpha}^{\dagger}(x)&=&\h{U}^{-1}(t,-\infty)\h{\psi}_{\alpha}^{(in)\dagger}(x)\h{U}(t,-\infty),
\\
\nonumber
\mbox{where} && \h{U}(t,-\infty) = \lim_{t_{0} \to -\infty} \h{U}(t,t_{0}) 
= \lim_{t_{0} \to -\infty}\h{T} e^{ \frac{1}{i \hbar} \int_{t_{0}}^{t} \d{t}^{\prime }\h{V}^{(in)}(t^{\prime})  },
\\
\nonumber
&&
\h{V}^{(in)}(t^{\prime}) 
\equiv  
\h{V}([\h{\psi}_{\alpha}^{(in)\,\dagger},\h{\psi}_{\alpha}^{(in)}];t^{\prime}) 
\eeqna
%
Combining \eq(def_psi_I) and \eq(j4_H), the four-element current density operator
in the interaction picture may be defined as:
$\h{j}^{(in)\,\mu}(x) = (c \h{\rho}^{(in)}(x),\,\hv{j}^{(in)}(x) )$.
These charge and current densities do not satisfy the charge conservation law,
except for $A=A^{(0)}$, and are merely convenient tools used for 
obtaining the expansion of the retarded product of the Heisenberg operators.
\beqna
\label{j4_HI}
\h{j}^{\mu}(x) \quad\;\,&=&\h{U}^{-1}(t,-\infty) \h{j}^{(in)\,\mu}(x) \h{U}(t,-\infty),
\\
\label{j4_I}
\h{j}^{(in)\,\mu}(x) &=&
\left\{
\begin{array}{lll}
c\, q \h{\psi}_{\alpha}^{(in)\dagger}(x)\, \h{\psi}_{\alpha}^{(in)}(x) & \text{ for }& \mu=0,
\\
\displaystyle
\h{\psi}_{\alpha}^{(in)\dagger}(x)  \frac{q}{2m}\left( \frac{\hbar}{i} (-\partial^{\mu}) -\frac{q}{c}A^{\mu}(x) \right) \h{\psi}_{\alpha}^{(in)}(x) + \mbox{h.c.} & \text{ for } & \mu = 1,2,3\,.
\end{array}
\right. 
\eeqna
To obtain the perturbative expansion (the retarded product series) of the Heisenberg operators,
let us introduce an operator in the intermediate picture, 
where $\h{U}(t,t_{0})$ will be used instead of  $\h{U}(t,-\infty)$:
\beqna
\nonumber
\h{\rho}^{\bullet}(x;t_{0}) 
\label{rho_C}
&=& \h{U}^{-1}(t, t_{0})  \,q\, \h{\psi}_{\alpha}^{(in)\dagger}(x)\, \h{\psi}_{\alpha}^{(in)}(x) \h{U}(t, t_{0})  ,
\\
\nonumber 
\h{j}_{i}^{\bullet}(x;t_{0})
\label{j_HI}
&=& \h{U}^{-1}(t, t_{0})  \frac{q}{2m}\h{\psi}_{\alpha}^{(in)\dagger}(x) \left( \frac{\hbar}{i} \partial_{i} -qA_{i}(x) \right) \h{\psi}_{\alpha}^{(in)}(x) \h{U}(t, t_{0}) + \mbox{h.c.} .
\eeqna
The corresponding four-element current density is 
\[
\h{j}^{\bullet\,\mu}(x;t_{0}) = (c\h{\rho}^{\bullet}(x;t_{0}),\, \hv{j}^{\bullet}(x;t_{0}) )
\]
As $t_{0} \to -\infty$, these operators coincide with those of the Heisenberg picture,
 while at $t_{0} = t$, they coincide with those of the interaction picture:
\beqna
\label{C2H}
\h{j}^{\bullet\,\mu}(x;-\infty)&=&\h{j}^{\mu}(x),
\\
\label{C2I}
\h{j}^{\bullet\, \mu}(x;t)\quad\;\,&=&\h{j}^{(in)\,\mu}(x).
\eeqna

Next, let's investigate the time evolution of $\h{j}^{\bullet\,\mu} $ as a function of $t_{0}$.
\beqna
\nonumber
\partial_{t_{0}} \h{j}^{\bullet\,\mu}(x;t_{0}) 
&=& 
\{ \partial_{t_{0}}  \h{U}^{-1}(t, t_{0}) \} \h{j}^{(in)\,\mu}(x) \h{U}(t, t_{0}) 
+
\h{U}^{-1}(t, t_{0}) \h{j}^{(in)\,\mu}(x) \{\partial_{t_{0}} \h{U}(t, t_{0})\} 
\\
\nonumber
&=&
\frac{1}{i\hbar} \h{V}^{(in)}(t_{0})   \h{U}^{-1}(t, t_{0}) \h{j}^{(in)\,\mu}(x)\h{U}(t, t_{0}) 
+
\h{U}^{-1}(t, t_{0}) \h{j}^{(in)\,\mu}(x) \h{U}(t, t_{0}) \frac{-1}{i\hbar} \h{V}^{(in)}(t_{0}) 
\\
\nonumber
&=&
\frac{-1}{i\hbar} \left[ \h{j}^{\bullet\,\mu}(x;t_{0}) ,   \h{V}^{(in)}(t_{0}) \right]
\eeqna
Integrating over $[t_{0},t]$, approximating iteratively using \eq(C2I), and 
changing the region of multi-integration, we obtain:
\beqna
\nonumber
\h{j}^{\bullet\,\mu}(x;t_{0}) 
&=& \h{j}^{(in)\,\mu}(x) + \frac{1}{i\hbar} \int_{t_{0}}^{t} \d{t}_{1}
 \left[ \h{j}^{\bullet\,\mu}(x;t_{1}) ,   \h{V}^{(in)}(t_{1}) \right]
\\
\nonumber
&=& \h{j}^{(in)\,\mu}(x) 
+ \frac{1}{i\hbar} \int_{t_{0}}^{t} \d{t}_{1} \left[ \h{j}^{(in)\,\mu}(x),   \h{V}^{(in)}(t_{1}) \right]
\\
\nonumber
&&\quad\quad
+ \left(\frac{1}{i\hbar}\right)^{2} \int_{t_{0}}^{t} \d{t}_{1} \int_{t_{1}}^{t} \d{t}_{2} 
\left[ \left[ \h{j}^{(in)\,\mu}(x),   \h{V}^{(in)}(t_{2}) \right],    \h{V}^{(in)}(t_{1}) \right]
 \\\
\nonumber
&&\quad\quad
+ \left(\frac{1}{i\hbar}\right)^{3} \int_{t_{0}}^{t} \d{t}_{1} \int_{t_{1}}^{t} \d{t}_{2}  \int_{t_{2}}^{t} \d{t}_{3} 
\left[ \left[ \left[ \h{j}^{(in)\,\mu}(x),   \h{V}^{(in)}(t_{3}) \right],  \h{V}^{(in)}(t_{2}) \right],    \h{V}^{(in)}(t_{1}) \right]+\cdots
\\
\nonumber
&=& \h{j}^{(in)\,\mu}(x) 
+ \frac{1}{i\hbar} \int_{t_{0}}^{t} \d{t}_{1} \left[ \h{j}^{(in)\,\mu}(x),   \h{V}^{(in)}(t_{1}) \right]
\\
\nonumber
&&\quad\quad
+ \left(\frac{1}{i\hbar}\right)^{2} \int_{t_{0}}^{t} \d{t}_{1} \int_{t_{0}}^{t_{1}} \d{t}_{2} 
\left[ \left[ \h{j}^{(in)\,\mu}(x),   \h{V}^{(in)}(t_{1}) \right],    \h{V}^{(in)}(t_{2}) \right]
 \\
\nonumber
&&\quad\quad
+ \left(\frac{1}{i\hbar}\right)^{3} \int_{t_{0}}^{t} \d{t}_{1} \int_{t_{0}}^{t_{1}} \d{t}_{2}  
\int_{t_{0}}^{t_{2}} \d{t}_{3} 
\left[ \left[ \left[ \h{j}^{(in)\,\mu}(x),   \h{V}^{(in)}(t_{1}) \right],  \h{V}^{(in)}(t_{2}) \right],    \h{V}^{(in)}(t_{3}) \right]+\cdots
\eeqna
Then, taking the limit $t_{0}\to-\infty$, the above equation yields 
the retarded product of the
Heisenberg operators, as follows:
\beqna
\nonumber
\h{j}^{\mu}(x) 
&=& \h{j}^{(in)\,\mu}(x) 
+ \frac{1}{i\hbar c} \int_{ct_{1}\in  (-\infty,ct]} \hspace{-2.0em}
\dn{4}{x}_{1} \left[ \h{j}^{(in)\,\mu}(x),   \h{v}^{(in)}(x_{1}) \right]
\\
\nonumber
&&
+ \left(\frac{1}{i\hbar c}\right)^{2} 
\int_{ct_{1}\in  (-\infty,ct]} \hspace{-2.0em} \dn{4}{x}_{1} 
\int_{ct_{2}\in  (-\infty,ct_{1}]} \hspace{-2.0em} \dn{4}{x}_{2} 
\left[ \left[ \h{j}^{(in)\,\mu}(x),   \h{v}^{(in)}(x_{1}) \right],    \h{v}^{(in)}(x_{2}) \right]
 \\
\nonumber
&&
+ \left(\frac{1}{i\hbar c}\right)^{3} 
\int_{ct_{1}\in (-\infty, ct]} \hspace{-2.0em} \dn{4}{x}_{1} 
\int_{ct_{2}\in (-\infty,ct_{1}]} \hspace{-2.0em} \dn{4}{x}_{2} 
\int_{ct_{3}\in (-\infty,ct_{2}]}\hspace{-2.0em} \dn{4}{x}_{3} 
\left[ \left[ \left[ \h{j}^{(in)\,\mu}(x),   \h{v}^{(in)}(x_{1}) \right],  \h{v}^{(in)}(x_{2}) \right],    \h{v}^{(in)}(x_{3}) \right]
\\
\label{RetProd}
&&
+\cdots
\\
\nonumber
\mbox{where}&& \h{V}^{(in)}(t) = \int \dn{3}{x} \,\h{v}^{(in)}(x)\,,
\\
\label{vin}
&& \h{v}^{(in)}(x) = 
\frac{1}{c}\Delta A^{\mu}(x)\, \h{j}^{(in0)}_{\mu}(x)  
-\frac{q}{2mc^{3}}\,
\t{\delta}_{\mu}^{\;\mu^{\prime}}
\Delta A^{\mu}(x) \Delta A_{\mu^{\prime}}(x)\, \h{j}^{(in0)}_{0}(x) \,,
\\
\label{jin}
&& \h{j}^{(in)\,\mu}(x) =  \h{j}^{(in0)\,\mu}(x) -\frac{q}{mc^{2}}\,
\t{\delta}_{\mu^{\prime}}^{\;\mu} 
\Delta A^{\mu^{\prime}}(x)\, \h{j}^{(in0)\,0}(x)\,,
\eeqna
and  $\h{j}^{(in0)\,\mu}(x)$ is the current density in the interaction picture,
that is, \eq(j4_I) with the VP being replaced by the non-perturbed system.
\Eq(vin) is obtained from \eq(Vdensity), 
replacing $\h{\psi}_{\alpha},\:\h{\psi}_{\alpha}^{\dagger}$ 
by $\h{\psi}_{\alpha}^{(in)},\:\h{\psi}_{\alpha}^{(in)\,\dagger}$, respectively.
Next, let us derive the single susceptibility in the form of Heisenberg operator 
by the functional derivative of \eq(RetProd) with respect to the EM potential.
 In \Eq(RetProd), the dependence of the EM potential through   
$\h{j}^{(in)\,\mu}(x)$ in \eq(j4_I) is of zeroth and first order for $\mu\in\{1,2,3\}$,
and dependence through $\h{v}^{(in)}(x_{1})$ is of first and second order.
The linear single susceptibility operator comes from 
the $A^{1}$-dependence, which exists in the first and second terms of \eq(RetProd) :
\beqna
\nonumber
\h{\chi}^{\mu}_{\;\;\mu_{1}}(x,x_{1}) 
&=&  
\left. \frac{ \delta \h{j}^{\mu}(x) }{ \delta A^{\mu_{1}}(x_{1} )}\right|_{A=A^{(0)}}
\\
\label{sus01}
=&&\hspace{-1.5em}
\frac{-q}{mc^{2}}\, \t{\delta}^{\mu}_{\;\;\mu_{1}} \delta^{4}(x-x_{1})\, \h{j}^{(in0)\,0}(x)
+\frac{1}{i\hbar c^{2} } \theta(ct-ct_{1}) \left[  \h{j}^{(in0)\,\mu}(x), \h{j}^{(in0)}_{\quad\mu_{1}}(x_{1}) \right],
\\
\nonumber
\mbox{where}&&
\h{j}^{(in0)\,\mu}(x)=\left. \h{j}^{(in)\,\mu}(x)\right|_{A=A^{(0)}}.
\eeqna

The Heisenberg operators of the nonlinear single susceptibilities, 
to second and higher order, are as follows.
To avoid any confusion in the case of two times coinciding,
the long and explicit expressions are given, without using the time ordering operator. 
\beqna
\nonumber
&& 
\hspace*{-8.0em}
2!\,
\h{\chi}^{\mu}_{\;\;\mu_{1}\mu_{2}}(x,x_{1},x_{2})
=
\left.
\frac{ \delta^{2} \h{j}^{\mu}(x) }
{ \delta A^{\mu_{1}}(x_{1} ) \delta A^{\mu_{2}}(x_{2}) } \right|_{A=A^{(0)}}
\\
\nonumber
=\frac{1}{i\hbar c ^{2}}\, \frac{-q}{mc^{2}}
&&
\left\{ 
\delta(ct-ct_{1}) \theta(ct-ct_{2})\quad
\t{\delta}^{\mu}_{\;\;\mu_{1}} \delta^{3}(x-x_{1}) 
\left[ \h{j}^{(in0)\,0}(x), \h{j}^{(in0)}_{\quad\mu_{2}}(x_{2}) \right] 
\right.
\\
\nonumber
&&+ 
\delta(ct-ct_{2}) \theta(ct-ct_{1})\quad
\t{\delta}^{\mu}_{\;\;\mu_{2}} \delta^{3}(x-x_{2}) 
\left[ \h{j}^{(in0)\,0}(x), \h{j}^{(in0)}_{\quad\mu_{1}}(x_{1}) \right] 
\\
\nonumber
&&\!\!\!+
\left.
\theta(ct-ct_{1})\delta(ct_{1}-ct_{2})
\t{\delta}_{\mu_{1}\,\mu_{2}} \delta^{3}(x_{1}-x_{2}) 
\left[ \h{j}^{(in0)\,\mu}(x), \h{j}^{(in0)}_{\quad 0}(x_{1}) \right] 
\right\}
\\
\nonumber
+\left(\frac{1}{i\hbar c^{2} }\right)^{2} 
&&
\left\{
\theta(ct-ct_{1}) \theta(ct_{1}-ct_{2})
\left[ \left[ \h{j}^{(in0)\,\mu}(x), \h{j}^{(in0)}_{\quad \mu_{1}}(x_{1}) \right], 
\h{j}^{(in0)}_{\quad \mu_{2}}(x_{2}) \right]   
\right.
\\
\label{sus02}
&&\!\!+ 
\left.
\theta(ct-ct_{2}) \theta(ct_{2}-ct_{1})
\left[ \left[ \h{j}^{(in0)\,\mu}(x), \h{j}^{(in0)}_{\quad \mu_{2}}(x_{2}) \right], 
\h{j}^{(in0)}_{\quad \mu_{1}}(x_{1}) \right]   
\right\} .
\eeqna
%
%
\beqna
\nonumber
&& 
3!\,
\h{\chi}^{\mu}_{\;\;\mu_{1}\mu_{2}\mu_{3}}(x,x_{1},x_{2},x_{3}) 
=  
\left.
\frac{ \delta^{3} \h{j}^{\mu}(x) }
{ \delta A^{\mu_{1}}(x_{1} ) \delta A^{\mu_{2}}(x_{2}) \delta A^{\mu_{3}}(x_{3}) } \right|_{A=A^{(0)}}
\\
\nonumber
&=&
\frac{1}{i\hbar c^{2} }  \left(\frac{-q}{mc^{2}}\right)^{2} 
\\
\nonumber
&&\left\{
\theta(ct-ct_{2}) \delta(ct-ct_{1}) \delta(ct_{2}-ct_{3}) 
\t{\delta}^{\mu}_{\;\;\mu_{1}} \delta^{3}(x-x_{1}) 
\t{\delta}_{\mu_{2}\,\mu_{3}} \delta^{3}(x_{2}-x_{3}) 
\left[ \h{j}^{(in0)\,0}(x), \h{j}^{(in0)}_{\quad\;\; 0}(x_{2}) \right] 
\right.
\\
\nonumber
&&+
\theta(ct-ct_{3}) \delta(ct-ct_{2}) \delta(ct_{3}-ct_{1}) 
\t{\delta}^{\mu}_{\;\;\mu_{2}} \delta^{3}(x-x_{2}) 
\t{\delta}_{\mu_{3}\,\mu_{1}} \delta^{3}(x_{3}-x_{1}) 
\left[ \h{j}^{(in0)\,0}(x), \h{j}^{(in0)}_{\quad\;\; 0}(x_{3}) \right] 
\\
\nonumber
&&+
\left.
\theta(ct-ct_{1}) \delta(ct-ct_{3}) \delta(ct_{1}-ct_{2}) 
\t{\delta}^{\mu}_{\;\;\mu_{3}} \delta^{3}(x-x_{3}) 
\t{\delta}_{\mu_{1}\,\mu_{2}} \delta^{3}(x_{1}-x_{2}) 
\left[ \h{j}^{(in0)\,0}(x), \h{j}^{(in0)}_{\quad\;\; 0}(x_{1}) \right] 
\right\}
\\
\nonumber
&&+
\left(\frac{1}{i\hbar c^{2}}\right)^{2} \frac{-q}{mc^{2}}
\\
\nonumber
&&
\left\{
\delta(ct-ct_{1})
\theta(ct_{1}-ct_{2}) 
\theta(ct_{2}-ct_{3})
\t{\delta}^{\mu}_{\;\;\mu_{1}}\delta^{3}(x-x_{1})
\left[ \left[ \h{j}^{(in0)\,0}(x), \h{j}^{(in0)}_{\quad \mu_{2}}(x_{2}) \right], 
\h{j}^{(in0)}_{\quad \mu_{3}}(x_{3}) \right]  
\right.
\\
\nonumber
&&+
\delta(ct-ct_{1})
\theta(ct_{1}-ct_{3}) 
\theta(ct_{3}-ct_{2})
\t{\delta}^{\mu}_{\;\;\mu_{1}}\delta^{3}(x-x_{1})
\left[ \left[ \h{j}^{(in0)\,0}(x), \h{j}^{(in0)}_{\quad \mu_{3}}(x_{3}) \right], 
\h{j}^{(in0)}_{\quad \mu_{2}}(x_{2}) \right]  
\\
\nonumber
&&+
\delta(ct-ct_{2})
\theta(ct_{2}-ct_{3}) 
\theta(ct_{3}-ct_{1})
\t{\delta}^{\mu}_{\;\;\mu_{2}}\delta^{3}(x-x_{2})
\left[ \left[ \h{j}^{(in0)\,0}(x), \h{j}^{(in0)}_{\quad \mu_{3}}(x_{3}) \right], 
\h{j}^{(in0)}_{\quad \mu_{1}}(x_{1}) \right]  
\\
\nonumber
&&+
\delta(ct-ct_{2})
\theta(ct_{2}-ct_{1}) 
\theta(ct_{1}-ct_{3})
\t{\delta}^{\mu}_{\;\;\mu_{2}}\delta^{3}(x-x_{2})
\left[ \left[ \h{j}^{(in0)\,0}(x), \h{j}^{(in0)}_{\quad \mu_{1}}(x_{1}) \right], 
\h{j}^{(in0)}_{\quad \mu_{3}}(x_{3}) \right]  
\\
\nonumber
&&+
\delta(ct-ct_{3})
\theta(ct_{3}-ct_{1}) 
\theta(ct_{1}-ct_{2})
\t{\delta}^{\mu}_{\;\;\mu_{3}}\delta^{3}(x-x_{3})
\left[ \left[ \h{j}^{(in0)\,0}(x), \h{j}^{(in0)}_{\quad \mu_{1}}(x_{1}) \right], 
\h{j}^{(in0)}_{\quad \mu_{2}}(x_{2}) \right]  
\\
\nonumber
&&+
\delta(ct-ct_{3})
\theta(ct_{3}-ct_{2}) 
\theta(ct_{2}-ct_{1})
\t{\delta}^{\mu}_{\;\;\mu_{3}}\delta^{3}(x-x_{3})
\left[ \left[ \h{j}^{(in0)\,0}(x), \h{j}^{(in0)}_{\quad \mu_{2}}(x_{2}) \right], 
\h{j}^{(in0)}_{\quad \mu_{1}}(x_{1}) \right]  
\\
\nonumber
&&+
\theta(ct-ct_{1}) \delta(ct_{1}-ct_{2})\theta(ct_{2}-ct_{3})
\t{\delta}_{\mu_{1}\,\mu_{2}}\delta^{3}(x_{1}-x_{2})
\left[ \left[ \h{j}^{(in0)\,\mu}(x), \h{j}^{(in0)}_{\quad 0}(x_{1}) \right], 
\h{j}^{(in0)}_{\quad \mu_{3}}(x_{3}) \right]  
\\
\nonumber
&&+
\theta(ct-ct_{2}) \delta(ct_{2}-ct_{3})\theta(ct_{3}-ct_{1})
\t{\delta}_{\mu_{2}\,\mu_{3}}\delta^{3}(x_{2}-x_{3})
\left[ \left[ \h{j}^{(in0)\,\mu}(x), \h{j}^{(in0)}_{\quad 0}(x_{2}) \right], 
\h{j}^{(in0)}_{\quad \mu_{1}}(x_{1}) \right]  
\\
\nonumber
&&+
\theta(ct-ct_{3}) \delta(ct_{3}-ct_{1})\theta(ct_{1}-ct_{2})
\t{\delta}_{\mu_{3}\,\mu_{1}}\delta^{3}(x_{3}-x_{1})
\left[ \left[ \h{j}^{(in0)\,\mu}(x), \h{j}^{(in0)}_{\quad 0}(x_{3}) \right], 
\h{j}^{(in0)}_{\quad \mu_{2}}(x_{2}) \right]  
\\
\nonumber
&&+
\theta(ct-ct_{1}) \theta(ct_{1}-ct_{2})\delta(ct_{2}-ct_{3})
\t{\delta}_{\mu_{2}\,\mu_{3}}\delta^{3}(x_{2}-x_{3})
\left[ \left[ \h{j}^{(in0)\,\mu}(x), \h{j}^{(in0)}_{\quad \mu_{1}}(x_{1}) \right], 
\h{j}^{(in0)}_{\quad 0}(x_{2}) \right]
\\
\nonumber
&&+
\theta(ct-ct_{2}) \theta(ct_{2}-ct_{3})\delta(ct_{3}-ct_{1})
\t{\delta}_{\mu_{3}\,\mu_{1}}\delta^{3}(x_{3}-x_{1})
\left[ \left[ \h{j}^{(in0)\,\mu}(x), \h{j}^{(in0)}_{\quad \mu_{2}}(x_{2}) \right], 
\h{j}^{(in0)}_{\quad 0}(x_{3}) \right]
\\
\nonumber
&&\left.+
\theta(ct-ct_{3}) \theta(ct_{3}-ct_{1})\delta(ct_{1}-ct_{2})
\t{\delta}_{\mu_{1}\,\mu_{2}}\delta^{3}(x_{1}-x_{2})
\left[ \left[ \h{j}^{(in0)\,\mu}(x), \h{j}^{(in0)}_{\quad \mu_{3}}(x_{3}) \right], 
\h{j}^{(in0)}_{\quad 0}(x_{1}) \right]
\right\}
\\
\nonumber
&&+
\left(\frac{1}{i\hbar c^{2} }\right)^{3}
\\
\nonumber
&&
\left\{ 
\theta(ct-ct_{1}) \theta(ct_{1}-ct_{2})  \theta(ct_{2}-ct_{3})
\left[ \left[ \left[ \h{j}^{(in0)\,\mu}(x), \h{j}^{(in0)}_{\quad \mu_{1}}(x_{1}) \right], 
\h{j}^{(in0)}_{\quad \mu_{2}}(x_{2}) \right] ,
\h{j}^{(in0)}_{\quad \mu_{3}}(x_{3}) \right]
\right.
\\
\nonumber
&&+
\theta(ct-ct_{1}) \theta(ct_{1}-ct_{3})  \theta(ct_{3}-ct_{2})
\left[ \left[ \left[ \h{j}^{(in0)\,\mu}(x), \h{j}^{(in0)}_{\quad \mu_{1}}(x_{1}) \right], 
\h{j}^{(in0)}_{\quad \mu_{3}}(x_{3}) \right] ,
\h{j}^{(in0)}_{\quad \mu_{2}}(x_{2}) \right]
\\
\nonumber
&&+
\theta(ct-ct_{2}) \theta(ct_{2}-ct_{3})  \theta(ct_{3}-ct_{1})
\left[ \left[ \left[ \h{j}^{(in0)\,\mu}(x), \h{j}^{(in0)}_{\quad \mu_{2}}(x_{2}) \right], 
\h{j}^{(in0)}_{\quad \mu_{3}}(x_{3}) \right] ,
\h{j}^{(in0)}_{\quad \mu_{1}}(x_{1}) \right]
\\
\nonumber
&&+
\theta(ct-ct_{2}) \theta(ct_{2}-ct_{1})  \theta(ct_{1}-ct_{3})
\left[ \left[ \left[ \h{j}^{(in0)\,\mu}(x), \h{j}^{(in0)}_{\quad \mu_{2}}(x_{2}) \right], 
\h{j}^{(in0)}_{\quad \mu_{1}}(x_{1}) \right] ,
\h{j}^{(in0)}_{\quad \mu_{3}}(x_{3}) \right]
\\
\nonumber
&&+
\theta(ct-ct_{3}) \theta(ct_{3}-ct_{1})  \theta(ct_{1}-ct_{2})
\left[ \left[ \left[ \h{j}^{(in0)\,\mu}(x), \h{j}^{(in0)}_{\quad \mu_{3}}(x_{3}) \right], 
\h{j}^{(in0)}_{\quad \mu_{1}}(x_{1}) \right] ,
\h{j}^{(in0)}_{\quad \mu_{2}}(x_{2}) \right]
\\
\label{sus03}
&&\left.+
\theta(ct-ct_{3}) \theta(ct_{3}-ct_{2})  \theta(ct_{2}-ct_{1})
\left[ \left[ \left[ \h{j}^{(in0)\,\mu}(x), \h{j}^{(in0)}_{\quad \mu_{3}}(x_{3}) \right], 
\h{j}^{(in0)}_{\quad \mu_{2}}(x_{2}) \right] ,
\h{j}^{(in0)}_{\quad \mu_{1}}(x_{1}) \right]
\right\}.
\eeqna
%
The charge conservation, \eq(ccl) and gauge invariance, \eq(gi) are respected in
\eqss(sus01)-(sus03). This fact is successfully checked after long and tedious calculations;
a supplementary document is provided for details.
%
\section{The Ground State in Density Functional Theory and Single Susceptibility}
\label{sec:eigenstate}
%
%
The linear and nonlinear single susceptibilities are the expectation values of 
the corresponding operators, \eqss(sus01)-(sus03), using the 
ground state in the non-perturbed electron system, 
which is specified by the simplified conditions in this paper:
\beq
\label{npSystem}
  \v{A}(x)=\v{A}^{(0)}(x) =\v{0} , \quad \v{j}^{\text{(EXT)}}(x) =\v{0},\quad
  \phi(x)=\phi^{(0)}(x)\;\text{and}\;\rho^{\text{(EXT)}}(x)\quad\text{are static.}
 \eeq
Let us explain how density functional theory\cite{DFT,Kohn-Sham} may allow us to prepare
the ground state and the complete set of the states in a many-electron system, 
refining the naive idea in Ref. \cite{SpringerFallacy}.
For that purpose, we need the electron field operators together with the SP and VP 
satisfying the coupled equations, \eqss(psi_opt)-(current00).
However, in the semiclassical treatment of the present theory, 
\eqs(charge00)-(current00) are replaced with 
their expectation values using the ground state, which we seek now on.
Due to this procedure, the quantum many-electron effect, the so-called exchange-correlation 
effect is ignored. Therefore, the solution of \eqss(psi_opt)-(current00) as it is
may not reproduce the electron charge density of the proper ground state, $\rho_{\text{GS}}(\v{r})$,
which is obtained using the ordinary Hamiltonian including 
the two-body Coulomb interaction, converted from the SP under the Coulomb gauge.
Such the electron density $\rho_{\text{GS}}(\v{r})$, in turn,  brings about the proper SP 
$\phi^{(0)}(x)$ under the Coulomb gauge.
Suppose that the proper electron charge density $\rho_{\text{GS}}(\v{r})$ is already known 
under the ordinary Hamiltonian.

Now, we like to seek for the ground state $| 0 \rangle$ in need,
adjusting the auxiliary potential $v^{\text{(AUX)}}(\v{r})$ to make
the electron charge density fit the proper one:
\beq
\label{KS}
\langle 0 | \h{\rho}(x) | 0 \rangle = \rho_{\text{GS}}(\v{r}).
\eeq
Such a situation in \eq(KS) is assumed by Kohn and Sham in 
the density functional theory\cite{Kohn-Sham}.
That is, \eqs(psi_opt)-(psiDagger_opt) are equivalent to Eq.(2.8) in Ref.\cite{Kohn-Sham}
[the Kohn-Sham equation], if  $v^{\text{(AUX)}}(\v{r}) $ is regarded as 
the so-called exchange-correlation potential.

For details, one may prepare the spin-orbital function $\varphi_{k}(\v{r})$
($k, \alpha$ stands for the orbital and spin states) 
as the eigenstate of the Kohn-Sham equation with the eigenenergy $\hbar \omega_{k}$.
Under the conditions of \eq(npSystem), the Kohn-Sham equation is, 
\begin{align}
0= \left( \hbar \omega_{k}   -q\phi^{(0)}(\v{r})    - \frac{1}{2m} 
\frac{\hbar}{i} \del{i} \cdot \frac{\hbar}{i} \del{i} -v^{\text{(AUX)}}(\v{r}) \right) \varphi_{k}(\v{r}),
\end{align}
where $v^{\text{(AUX)}}(\v{r}) $ is set to the exchange-correlation potential, that 
guarantees \eq(KS).
Then, $\h{\psi}_{\alpha}(x) = \h{\psi}_{\alpha}^{(in0)}(x) =
\sum_{k} \varphi_{k}(\v{r})\,\h{a}_{k\alpha}^{(in0)}(t)$
satisfies \eq(psi_opt) under the condition \eq(npSystem), where $\h{a}_{k\alpha}^{(in0)}$ 
is the operator to annihilate the electron associated with 
the spin-orbital  $\varphi_{k}(\v{r})$ in the non-perturbative system.
Considering $\{ \varphi_{k}(\v{r}) \}$ as a complete set of the one-electron functional space,
the ground state with the electron number $n$
in the present theory is constructed as
the single Slater determinant,
\begin{align}
| 0 \rangle = \lim_{t_{0}\to -\infty}\,\frac{1}{\sqrt{n!}}\prod_{k \alpha} \h{a}^{(in0)\,\dagger}_{k\alpha}(t_{0})
\,|\text{vac}\rangle\,,
\end{align}
where $|\text{vac}\rangle$ is the vacuum state, and the indecies ${k \alpha}$ scan over 
the $n$ spin-orbitals from the lowest eigenenergies.   
Furthermore,  under the fixed $v^{\text{(AUX)}}(\v{r}) $ and $\phi^{(0)}(\v{r})$,
one may consider all the possible combination of $n$ spin-orbitals 
and obtain the normalized orthogonal complete set
 $\{ | m \rangle | m = 0,1,2,\cdots \}$ 
 in terms of all the possible single Slater determinants.

On the above logic, one should know the proper electron charge density $\rho_{\text{GS}}(\v{r})$ beforehand to determine  $v^{\text{(AUX)}}(\v{r})$,
which is the universal functional of the electron density\cite{DFT,Kohn-Sham}.
In practice, however, one may solve the Kohn-Sham equation, possibly
under the local density approximation 
for $v^{\text{(AUX)}}(\v{r})$,
and reconsider the resulting charge density as
$\rho_{\text{GS}}(\v{r})$.

The expectation value of the single susceptibility operator is,
$
\langle 0 | \h{\chi}^{\mu}_{\;\mu_{1}\cdots}(x, x_{1}, \cdots) | 0 \rangle
$,
and, for example, the linear susceptibility becomes:
\begin{align}
\nonumber
\langle 0 |  \h{\chi}^{\mu}_{\;\;\mu_{1}}(x,x_{1}) | 0 \rangle
=&
\frac{-q}{mc^{2}}\, \t{\delta}^{\mu}_{\;\;\mu_{1}} \delta^{4}(x-x_{1})\, \langle 0 | \h{j}^{(in0)\,0}(x) | 0 \rangle
\\
\label{sus01_expectation}
&+\frac{1}{i\hbar c^{2} } \theta(ct-ct_{1}) \langle 0 |  \left[  \h{j}^{(in0)\,\mu}(x), \h{j}^{(in0)}_{\quad\mu_{1}}(x_{1}) \right] | 0 \rangle\,.
\end{align}
Next, to evaluate the products of two (or more) current density operators,
e.g., the second term in \eq(sus01_expectation), 
we may use the projection operator 
$\h{1}=\sum_{m} | m \rangle \langle m |$.
Now, the expectation value in the second term of \eq(sus01_expectation) becomes,
\begin{align}
\nonumber
 & \langle 0 |  \left[  \h{j}^{(in0)\,\mu}(x), \h{j}^{(in0)}_{\quad\mu_{1}}(x_{1}) \right] | 0 \rangle
 \\
 \nonumber
 & = \sum_{m} 
 \left\{
\langle 0 |  \h{j}^{(in0)\,\mu}(x)   | m \rangle 
\langle m | \h{j}^{(in0)}_{\quad\mu_{1}}(x_{1}) | 0 \rangle 
  -
\langle 0 |  \h{j}^{(in0)}_{\quad\mu_{1}}(x_{1})  | m \rangle
\langle m |  \h{j}^{(in0)\,\mu}(x)  | 0 \rangle 
\right\}
 \\
 \nonumber
 & = \sum_{m} 
 \lim_{t_{0} \to \infty} \left\{
\langle 0 |  e^{\frac{-1}{i\hbar}\h{H}^{(0)}(t-t_{0} )} \h{j}^{(in0)\,\mu}(x)|_{t=t_{0}} e^{\frac{1}{i\hbar}\h{H}^{(0)}(t-t_{0} )}  | m \rangle 
\langle m | e^{\frac{-1}{i\hbar}\h{H}^{(0)}(t_{1}-t_{0} )} \h{j}^{(in0)}_{\quad\mu_{1}}(x_{1}) |_{t_{1}=t_{0}} e^{\frac{1}{i\hbar}\h{H}^{(0)}(t_{1}-t_{0} )} | 0 \rangle 
\right.
\\
\nonumber
 &\quad\quad\quad\quad\;\,-
\left.
\langle 0 | e^{\frac{-1}{i\hbar}\h{H}^{(0)}(t_{1}-t_{0} )}  \h{j}^{(in0)}_{\quad\mu_{1}}(x_{1})|_{t_{1}=t_{0}} e^{\frac{1}{i\hbar}\h{H}^{(0)}(t_{1}-t_{0}) } | m \rangle
\langle m | e^{\frac{-1}{i\hbar}\h{H}^{(0)}(t-t_{0} )}  \h{j}^{(in0)\,\mu}(x)|_{t=t_{0}} e^{\frac{1}{i\hbar}\h{H}^{(0)}(t-t_{0}) } | 0 \rangle 
\right\}
 \\
 \nonumber
 & = \sum_{m} 
\left\{
e^{\frac{1}{i\hbar}(E_{m}-E_{0})(t-t_{1} )}
\langle 0|  \h{j}^{(in0)\,\mu}(x)|_{t=-\infty} | m \rangle 
\langle m |  \h{j}^{(in0)}_{\quad\mu_{1}}(x_{1}) |_{t_{1}=-\infty} | 0 \rangle 
\right.
\\
\label{representation}
 &\quad\quad\quad\,-
\left.
e^{\frac{-1}{i\hbar}(E_{m}-E_{0})(t-t_{1} )}
\langle 0 | \h{j}^{(in0)}_{\quad\mu_{1}}(x_{1})|_{t_{1}=-\infty}  | m \rangle
\langle m |  \h{j}^{(in0)\,\mu}(x)|_{t=-\infty} | 0 \rangle 
\right\}
.
\end{align}
In the induced charge and current densities obtained from
the convolution integral of \eq(representation) with the perturbative EM field,
the energy denominator will appear
as shown in \S\ref{sec:oneElectron}.

In the above theoretical framework, 
$ |m \rangle $'s are simply the members of the complete set, 
and, in principle,  do not carry physical meaning of excited states of  
a many-electron system. Considering that 
the density functional theory concerns only the ground state of 
the many-electron system, 
the above treatment is a sound application of density functional theory 
to the response theory adequate for NFO.
Remark that the variational principle based on \eq(action) cannot determine
the auxiliary potential, $v^{\text{(AUX)}}(x)$ but is determined with the help of
another theory, namely, the density functional theory.

As a summary, 
the quantum many-electron effect is temporally ignored in the present semiclassical theory,
but is compensated with the support of the density functional theory. 
In other words, the SP inherently existing in the electron system
is separated as $\phi^{(0)}(x)$ and $v^{\text{(AUX)}}(\v{r})$, and  
the SP incidence may be treated equally with the VP incidence.  
Note that, $\phi^{(0)}(x)$ is under the Coulomb gauge but the SP and VP incidences
may be gauge-free, that is,  the present response theory is still free from gauge-fixing.
%
%
%
%
\section{Application: A Logical Fallacy to use the Electric Field in Near-field Optics}
\label{sec:oneElectron}
%
Under {\it non-resonant conditions} in the optical near field,
{\it non-metallic materials} cause various phenomena not 
observed in conventional optics, such as 
highly efficient light emission from 
indirect-transition-type semiconductors (LED\cite{SiLED,OhtsuSiLED} and 
Laser\cite{SiLaser,OhtsuSiLED}),
chemical reaction with insufficient photon energy 
(chemical vapor deposition\cite{NFCVD},  optical NF lithography\cite{NFLitho}, 
optical NF etching\cite{NFEtching}),
frequency up-conversion\cite{NFUpConv,PlsShpMs},
non-adiabatic effect beyond forbidden transition 
(local energy concentration\cite{nanoFountain},
nano-photonic gate device\cite{nanoPhotonicDevice}),
and gigantic magneto-optical rotation of the LED\cite{OhtsuSiLED,MO}.

These experimental results draw attention to a fundamental role of the non-resonant condition 
in NFO.
We have no complete answer at this stage 
but the application of the present response theory to a many-electron system in NFO
shows a logical fallacy to use the electric field and the electric permittivity,
and the necessity to use the EM potential and the associated single susceptibility.
The discussion of the {\it one-electron} system appeared in Ref.\cite{SpringerFallacy},
but is concisely reviewed below in \S\ref{sec:Elt}-\S\ref{sec:NF} because 
the {\it many-electron} version in \S\ref{sec:MEextension} may be simply
a recast of the one-electron version, owing to 
the density functional theory.
For the readability, calculation details are given in Appendix \ref{sec:methods}.
%
 \subsection{Classification of optical systems}
\label{sec:classification}
%
First, let us classify the optical systems. The two systems under near- and far-field incidence conditions 
in \fig{fig:NFOvsOO} are subdivided into two classes
depending on the near- or far-field observation condition.
These four classes are listed in TABLE \ref{table:NFOvsOO},  
together with a summary of the results mentioned below.
In particular, 
the systems of (I$^{\prime}$) and (I\hspace{-.1em}I$^{\prime}$)
are the limiting cases of null longitudinal incidence of 
the systems (I) and (I\hspace{-.1em}I), respectively.
Thus, in the systems (I$^{\prime}$) and (I\hspace{-.1em}I$^{\prime}$),
the longitudinal response vanishes and 
the difference in response may not be observed.
In the following, therefore, we focus mainly on 
systems (I) and (I\hspace{-.1em}I),
in which longitudinal incidence exists.
\begin{table}[tb]
\caption{
Classification of optical systems by distance 
from the target material to the light source
and distance from that to the observation point, together with 
a summary of the results; the validity of the electric field as the cause of the response. 
}
\label{table:NFOvsOO}
\begin{tabular}{| l | l | l |}\hline\hline
	                      & \bf Near-field observation   &\bf Far-field observation\\
	                      & \quad     Source:$\Delta\rho$ and $ \Delta \v{j}$ & \quad     Source: $\overline{\Delta \v{j}}$ \\	                     \hline\hline
$\displaystyle \begin{matrix*}[l]
\text{\bf Near-field incidence :}\\
\quad\Delta\v{E}^{(\ell)}+\Delta\v{E}^{(t)}
\end{matrix*}$
& \ovalbox{$\displaystyle \begin{matrix*}[l]
\text{\bf (I) NF optical system}\\
\text{non-resonant / resonant}
\end{matrix*}$}
&  \ovalbox{$\displaystyle \begin{matrix*}[l]
\text{\bf (I\hspace{-.1em}I) NF optical system}\\
\text{non-resonant / resonant}
\end{matrix*}$}
\\
\quad Validity of the electric field 
&\quad\quad\quad\quad \colorbox{black}{\textcolor{white}{NG}} / OK & \;\quad\quad\quad\quad OK / OK 
\\\hline
$\displaystyle \begin{matrix*}[l]
\text{\bf Far-field incidence :}\\
\quad \Delta\v{E}^{(t)}
\end{matrix*}$
& \ovalbox{$\displaystyle \begin{matrix*}[l]
\text{\bf (I$^{\prime}$) NF optical system}\\ 
\text{non-resonant / resonant}
\end{matrix*}$}
& \ovalbox{$\displaystyle \begin{matrix*}[l]
\text{\bf (I\hspace{-.1em}I$^{\prime}$\!){\footnotesize conventional optical\! system}}\!\\
\text{non-resonant / resonant}
\end{matrix*}$}
\\
\quad Validity of the  electric field
&\;\quad\quad\quad\quad OK / OK & \;\quad\quad\quad\quad OK / OK 
\\\hline\hline
\end{tabular}
\end{table}
%
 \subsection{Susceptibilities associated with longitudinal and transverse electric fields}
\label{sec:Elt}
%
Applying the present linear response theory and the long wave approximation (LWA)
to the spinless one-electron system with two levels on a small scale,  
the induced charge and current densities ({\it as a result of the response}), 
$\Delta\rho(\v{r},t)$ and $\Delta\v{j}(\v{r},t)$,
are described as the total derivative with respect to
the longitudinal and transverse electric fields ({\it as the cause of the response}),
$\Delta \v{E}^{(\ell)}(\v{0},t)$ and $\Delta \v{E}^{(t)}(\v{0},t)$, 
where $\v{0}$ is the representative position in the electron system under the LWA.
The derivations are given in \S\ref{cd:1} and the results are, 
\begin{align}
\label{rho_ind_av}
\Delta\rho (\v{r},t)
&= 
\chi^{\rho \leftarrow (\ell)}_{j}(\v{r},\omega) \,\Delta E^{(\ell)}_{j}(\v{0},t)\,
+
\chi^{\rho \leftarrow(t)}_{j}(\v{r},\omega) \,\Delta E^{(t)}_{j}(\v{0},t)\,,
\\
\label{j_ind_av}
\Delta j_{i} (\v{r},t)
&= 
\chi^{\v{j} \leftarrow (\ell)}_{ij}(\v{r},\omega)\,\Delta \dot{E}^{(\ell)}_{j}(\v{0},t)
+
\chi^{\v{j} \leftarrow (t)}_{ij}(\v{r},\omega)\,\Delta \dot{E}^{(t)}_{j}(\v{0},t)\,,
\end{align}
where the partial derivative coefficients, $\chi^{\cdots}_{\cdots}(\v{r},\omega)$'s
are susceptibilities associated with the longitudinal and transversal electric fields.
In \eq(j_ind_av), the time derivatives of the two types of electric fields, namely,
$\Delta \dot{E}^{(\ell)}_{j}(\v{0},t)$ and $\Delta \dot{E}^{(t)}_{j}(\v{0},t)$,
are regarded as the causes, instead of the two types of electric fields themselves.
The magnetic response will appear in the higher order of the LWA and
vanishes in \eqs(rho_ind_av)-(j_ind_av) representing  the leading order; 
see Refs.\cite{Cho01,Cho02} and \S\ref{sec:existing}.
For the present spinless electron system, the electron field operators,
$\h{\psi}_{\alpha}^{\dagger}(x),\, \h{\psi}_{\alpha}(x)$ is reconsidered as 
$\h{\psi}^{\dagger}(x),\, \h{\psi}(x)$, respectively, eliminating the index of the spin state, ${\alpha}$. 

To evaluate the susceptibilities in \eqs(rho_ind_av)-(j_ind_av),
the two levels are assumed to be the ground and excited states in the non-perturbed system
with eigenenergies, $\hbar\omega_{0}$ and $\hbar\omega_{1}$, 
and orbitals, $\varphi_{0}(\v{r})$ and $\varphi_{1}(\v{r})$, respectively.
Those orbitals are assumed to be bound states expressed by real functions,
carry well-defined and distinct spatial parities (even and odd parities), 
and form the normalized orthogonal complete set.
The excitation energy is 
$\hbar\Delta\omega_{1}\equiv\hbar\omega_{1}-\hbar\omega_{0} \,>\,0$;
this finite excitation energy means that the target is a non-metallic material, such as
a molecule, nano-structured semiconductor and insulator. 

The susceptibilities in \eqs(rho_ind_av)-(j_ind_av) are derived in \S\ref{cd:1},
and those leading to the induced charge density result in the following:
\begin{align}
\label{suscept4rho01}
\chi^{\rho \leftarrow (\ell)}_{j}(\v{r},\omega) 
=&
\chi^{\rho \leftarrow (t)}_{j}(\v{r},\omega)
=
2q^{2} \, 
\frac{\eta}{\eta^{2}-1}\,
\frac{1}{\hbar\omega} 
\,\c{D}_{j}\,\varphi_{0}(\v{r}) \varphi_{1}(\v{r}) \,,
\\
\text{where} \quad\quad
\label{defeta}
\eta \equiv& \frac{\hbar \Delta\omega_{1}}{\hbar\omega} 
= \frac{\text{excitation energy}}{\text{photon energy}}\,,\;\text{and}
\\
\label{dtme}
\c{D}_{i}\equiv& \int \dn{3}{r} \: \varphi_{1}(\v{r}) \:  r_{i}\: \varphi_{0}(\v{r})\,.
\end{align}
This means that the responses to the longitudinal and transverse electric fields are common, 
such that the induced charge density has a linear relationship with
{\it the total electric field}, 
namely, 
$\displaystyle
\Delta\rho (\v{r},t) =
\chi^{\rho \leftarrow (\ell)\text{\:or\:}(t)}_{j}(\v{r},\omega) 
\left(
\Delta E^{(\ell)}_{j}(\v{0},t)+\Delta E^{(t)}_{j}(\v{0},t)
\right)
$.

The susceptibilities leading to the induced current density are not so simple and 
result in the following:
\begin{align}
\label{suscept4j0}
\chi^{\v{j} \leftarrow (\ell)}_{ij}(\v{r},\omega) 
=&
\frac{q^{2}\hbar^{2}\,}{m}\,
\frac{1}{\eta^{2}-1}\,
\frac{1} {(\hbar\omega)^{2}}\,
\c{D}_{j}\,
\left(    \partial_{i} \varphi_{1}(\v{r}) \varphi_{0}(\v{r}) - \varphi_{1}(\v{r}) \partial_{i} \varphi_{0}(\v{r}) 
\right)\,,
\\
\label{suscept4j1}
\chi^{\v{j} \leftarrow (t)}_{ij}(\v{r},\omega)
=&
\eta^{2}\,\chi^{\v{j} \leftarrow (\ell)}_{ij}(\v{r},\omega) 
- \frac{q^{2}\hbar^{2}\,}{m}\,
\frac{1} {(\hbar\omega)^{2}}\,
\varphi_{0}(\v{r})\varphi_{0}(\v{r})\,.
\end{align}
The susceptibility to the transverse electric field, \eq(suscept4j1), 
is composed of two terms.
The first term, namely, the resonant term, includes the energy denominator enhanced under the resonant condition, $\eta \simeq 1$, as in the susceptibility to the longitudinal electric field,
\eq(suscept4j0).
The second term, namely, the non-resonant term, does not include such a resonance factor.
%
\subsection{Equal responses under the resonant condition}
\label{sec:Res}
%
Under the condition $\eta\simeq 1$ in all cases in TABLE \ref{table:NFOvsOO},
\eq(suscept4j1) is dominated by the resonant term (the first term) 
over the non-resonant term (the second term)
and asymptotically equals \eq(suscept4j0).
\begin{align}
\label{suscept4j01}
\chi^{\v{j} \leftarrow (t)}_{ij}(\v{r},\omega)\simeq \chi^{\v{j} \leftarrow (\ell)}_{ij}(\v{r},\omega)\,.
\end{align} 
\Eq(suscept4j01) together with \eq(suscept4rho01) reveal the equivalency of the responses 
to the longitudinal and transverse electric fields, so that 
the total electric field is regarded as the cause of the response in all the optical systems
under the resonant condition 
listed in TABLE \ref{table:NFOvsOO}.
%
\subsection{Equal responses under the far-field observation condition}
\label{sec:FF}
%
In the system (I\hspace{-.1em}I) and (I\hspace{-.1em}I$^{\prime}$) in TABLE \ref{table:NFOvsOO},
the far field to be observed is insensitive to the details of the source but is determined by the
spatial average of the source.
Under the LWA, such an average can be achieved by the spatial average of the susceptibilities. 
Detailed calculations are shown in \S\ref{cd:3} and the results are as follows:
\begin{align}
\label{suscept4rho_far}
\overline{\chi^{\rho \leftarrow (\ell)}_{j}(\v{r},\omega)} 
&=
\quad
\overline{\chi^{\rho \leftarrow (t)}_{j}(\v{r},\omega)}  =0\,,
\\
\label{suscept4j_far}
\overline{\chi^{\v{j} \leftarrow (\ell)}_{ij}(\v{r},\omega)}
&=
\overline{\chi^{\v{j} \leftarrow (t)}_{ij}(\v{r},\omega)}
=
\delta_{i\,j} \frac{q^{2}\hbar^{2}}{m\,\c{V}}  
\frac{1} {(\hbar\Delta \omega_{1})^{\,2}-(\hbar\omega)^{2}}\,,
\end{align}
where the overline represents the spatial average and $\c{V}$ is the
volume of the target material. 
From \eqs(suscept4rho_far)-(suscept4j_far), one may not observe 
different responses to the two types of incidences under the far-field observation condition.
The null response represented in \eq(suscept4rho_far) 
is reasonable because the induced charge density yields the longitudinal electric field,
which has a non-radiative nature and vanishes in the far-field regime.

\subsection{Unequal responses under the non-resonant, NF incidence, and NF observation conditions}
\label{sec:NF}
%
The different responses to the longitudinal and transverse electric fields
claimed in \S\ref{sec:equalTreatment} may be detected only in the system (I) in TABLE \ref{table:NFOvsOO}
under the non-resonant condition, which is just 
the compliment to the popular optical systems
under the resonant condition or the far-field incidence condition or the far-field observation condition.
In the NF optical system (I) with a non-metallic material under the non-resonant condition, 
{\it the total electric field}  is not the cause of the response;
therefore, the response may not be described by 
the ordinary constitutive equation, namely,
the linear relationship between the polarization and "electric field" via the electric permittivity,
so that the single susceptibility is essential to treat separately 
the longitudinal and transverse incidences. 
\subsection{Extension to the many-electron system}
\label{sec:MEextension}
The above one-electron model is very simple and the responses 
may be modified in a many-electron system or a low-symmetry system. 
However, 
the difference in the responses to the two types of electric fields 
originates in {\it the non-relativistic nature} 
of the system (as stated in \S\ref{sec:equalTreatment}),
and should survive in actual NF optical systems with non-metallic materials
(the materials with finite excitation energy). 
Actually, the results revealed in \S\ref{sec:Elt}-\S\ref{sec:NF} are applicable to the 
corresponding many-electron system,
considering the auxiliary potential $ v^{\text{(AUX)}}(x) $  to construct the orbitals 
using the Kohn-Sham equation (\ref{KS}), and
replacing the complete orthogonal set composed of the one-electron ground and excited states, 
$ \{\h{a}_{0}^{(in0)\,\dagger}(-\infty)| \text{vac} \rangle,\, 
\h{a}_{1}^{(in0)\,\dagger}(-\infty) | \text{vac} \rangle \}$ ($-\infty$ means the time of the infinite past)
 to the corresponding set, composed of two single Slater determinants, 
 $\{ |0 \rangle,\, \h{a}_{1}^{(in0)\,\dagger}(-\infty) \h{a}_{0}^{(in0)}(-\infty) |0\rangle \}$, 
 where $|0 \rangle$ is the ground state in the density functional theory as defined in 
 \S\ref{sec:eigenstate}, and 
 $\h{a}_{0}^{(in0)}[,\,\h{a}_{0}^{(in0)\,\dagger}]$ and $\h{a}_{1}^{(in0)}[,\,\h{a}_{1}^{(in0)\,\dagger}]$
are the annihilation [and creation] operators associated by the 
 the highest occupied molecular orbital (HOMO) and 
 the lowest unoccupied molecular orbital (LUMO), respectively,
 determined by the Kohn-Sham equation (\ref{KS}).
Owing to the density functional theory, 
recasting the formulation in the one-electron system
brings that in the many-electron system, 
if the HOMO and LUMO dominate the excitation.

As a result, the many-electron system of a non-metallic material 
under the non-resonant, NF incidence, and NF observation conditions
may not be described in terms of the electric field and the associated permittivity.
Instead, the EM potential and the single susceptibility are essential.
%
\subsection{Comparison with the existing theories}
\label{sec:existing}
In NFO, the response to the longitudinal electric field
is discussed in Chap. 5 in Ref.\cite{Cho02} and Chap. 9 in Ref.\cite{Keller01},
as mentioned in \S\ref{sec:MEP}.
The present work is a further comparison of the responses
to the two-types of electric field, 
considering {\it the non-resonant condition}. 

Another logical fallacy to use the electric and magnetic fields
is pointed out by Cho, as briefly mentioned in \S\ref{sec:equalTreatment}.
In Refs.\cite{Cho01,Cho02},
Cho derived a Taylor series of the nonlocal response function\cite{Cho00} under the LWA,
and assigned the electric permittivity and magnetic permeability in the macroscopic constitutive equation
as the term of order $\c{O}(ka)^{0}$ (the leading order) and $\c{O}(ka)^{2}$, respectively, 
where $ka \ll1$, $2\pi/k$ is the light wavelength, and $a$ is the representative size of the material. 
Furthermore, he pointed out that 
the ordinary two susceptibilities are irrational because the separability of the electric and magnetic responses not applicable and the term of order $\c{O}(ka)^{1}$ appears in a chiral symmetric system, including a NF optical system with a low-symmetric nanostructure.
The present demonstration is concerned with the logical fallacy, which appears in the electric response (the leading order from the viewpoint of Cho) in NFO under {\it a non-resonant condition}.
%
\subsection{A remark on applying the finite differential time domain (FDTD) method to 
NF optical systems}
\label{sec:FDTD}
The macroscopic constitutive equations in terms of the electron permittivity
and magnetic permeability 
have been widely employed to calculate the optical near field 
in the FDTD method\cite{FDTD}.
One may notice that the permittivity in the FDTD method carries a simple spatial dependence 
and leads to some quantitative error.
Actually, the microscopic susceptibilities, for example, \eq(suscept4rho01), \eq(suscept4j0), and \eq(suscept4j1), have rippling spatial distributions 
originating from the orbitals.

In the case of the NF optical system (I) in TABLE \ref{table:NFOvsOO}
with a non-metallic material under the non-resonant condition,
the situation is more serious because 
the concept {\it electric field} is not available, such that
it is a logical fallacy to use the macroscopic constitutive equation. Thus, 
a novel simulation method is necessary, in particular, for the NF optical system
with a non-metallic material.
%
\subsection{Why this fallacy has been missed for a long time?}
%
Why has the comparison of responses to the two types of electric fields 
not been addressed in NF optical theory?
First, in the long history of optics, 
the NF optical system (I) in TABLE \ref{table:NFOvsOO} under a non-resonant condition 
has been out of focus. 
Such a system could not be resolved until 
the technical difficulty of NF observation was overcome. Additionally, resonance phenomena continue to
attract attention. Furthermore, 
even in NFO, there has been less emphasis on non-metallic materials, as opposed to metallic materials, which are essential for 
plasmonics.

The second reason is that the ordinary Hamiltonian for a many-electron system does not include the longitudinal electric field, which is rewritten to the two-body Coulomb interaction, as stated in \S\ref{sec:MEP}.
With this Hamiltonian, the non-linear response to the longitudinal electric field 
(the SP under the Coulomb gauge) incidence accompanies the Coulomb interaction,  
and is ignored or unequally treated 
compared with the response to the transverse electric field (the VP under the Coulomb gauge).
\subsection{Summary of this section}
In the NF optical system (I) in TABLE \ref{table:NFOvsOO},
the responses to the longitudinal and transverse electric fields
should be separately treated, and in a more general view point beyond the LWA and 
linear response theory,  it is essential to employ the linear and nonlinear single susceptibilities,  
considering both of the SP and VP equally as the cause of response.

To the best of our knowledge, 
the NF optical system with non-metallic material
under the non-resonant condition, namely, the system (I) in TABLE \ref{table:NFOvsOO},
is the third example that cannot be described in terms of the electric field and/or 
magnetic field, after the superconductor system with the
Meissner effect\cite{London} and the electron system with the Aharonov-Bohm effect\cite{ABeffect},
as mentioned in \S\ref{sec:singleSus}. 
%
%
\section{Summary}
\label{sec:summary}
%
\begin{enumerate}
\item
Aiming to investigate electron response in NFO, 
we define the linear and nonlinear single susceptibilities, 
equally considering the SP and VP as the cause of the response.
\item
It is shown that the present single linear and nonlinear susceptibilities 
guarantee charge conservation and gauge invariance.
\item
The linear and nonlinear susceptibilities in the form of Heisenberg operators
are derived systematically by means of the functional derivatives 
of the action integral of the matter with respect to the SP and VP.
\item
It is shown that the density functional theory may 
be used  in the non-perturbed system and support to prepare 
the ground state and a complete set of states, 
which in turn are used to evaluate the expectation values of the operators 
of the linear and nonlinear susceptibilities.
\item
Applying the present response theory to a simplified model system, 
it is shown that the single susceptibility is essential to
describe the response of the optical system with
non-metallic material under the non-resonant, NF incidence, and NF observation conditions.
\end{enumerate}
Some remaining problems meriting further investigation include:
\begin{enumerate}
\item
Applying the present response theory to actual non-resonant NF optical systems 
with a non-metallic material in Refs.\cite{SiLED}-\cite{MO} 
to explore the mechanism leading to the outstanding experimental results 
such as the high-efficient light emission and gigantic magneto-optical effect, etc.
\item
Developing a constitutive equation based on the single susceptibility
which can aid experimentalists in NFO
as a substitute for the electric permittivity and magnetic permeability of
ordinary optics.
\item
Developing a practical simulator for the many-electron system in NFO,
using the present response theory with the support of the density functional theory,
as the replacement of the FDTD simulation method, 
\item   
Extending the response theory to treat the spin-polarization system in NFO, 
based on the Pauli or Dirac equation. 
\end{enumerate}
\begin{acknowledgements}
The author thanks Prof. K. Cho in Osaka Univ. for useful discussion about the 
single susceptibility. He also thanks Prof. M. Ohtsu (Univ. of Tokyo, Research Origin of Dressed Photon (RODreP)) and the members of his group in Univ. of Tokyo, 
Prof. T. Kawazoe (Tokyo Denki Univ.) for the comment from the experimental view point
and dressed photons;   
Prof. I. Ojima (RODreP),  Prof. H. Saigo (Nagahama Insitute of Bio-Science and Technology), 
Drs H. Sakuma (RODreP), K. Okamura (Nagoya Univ.),  H. Ando (Chiba Univ.) 
for useful discussions on the context of dressed photon.
This work is partially supported by JSPS KAKENHI Grant Number JP25610071 during 2013-2015, 
Research Foundation for Opto-Science and Technology during 2018-2019, 
and Research Origin for Dressed Photons.
\end{acknowledgements}
%
%
\appendix
%
%
\section{Optimization of Electron Field Operators Under Arbitrary EM Potential}
\label{sec:derivativeZero}
%
Under a given EM potential, $A^{\nu}$, the electron field operator optimized to satisfy \eq(psi_opt)
is considered as the functional of $A^{\nu}$, i.e., $\h{\psi}_{\alpha}(x;[A^{\nu}]), \h{\psi}_{\alpha}^{\dagger}(x;[A^{\nu}])$.
Then, the next equation holds for $n=0,1,2,\cdots$:
\beqna
\label{nFunctionalDerivativeA01}
&&
\left. \frac{\delta^{n} }{\delta A^{\mu_{n}}(x_{n}) \cdots \delta A^{\mu_{1}}(x_{1})} 
\delta \h{\psi}_{\alpha}^{\dagger}(x^{\prime}) {\Large \backslash} \delta\c{I}_{\mbox{mat}} 
\,\right|_{A^{\nu}=A^{(0)\nu}} 
=0,
\\
\label{nFunctionalDerivativeA02}
&&
\left. \frac{\delta^{n} }{\delta A^{\mu_{n}}(x_{n}) \cdots \delta A^{\mu_{1}}(x_{1})} 
\delta\c{I}_{\mbox{mat}} / \delta \h{\psi}_{\alpha}(x^{\prime})
\,\right|_{A^{\nu}=A^{(0)\nu}}
=0.
\eeqna
Proof:
\Eq(psi_opt) should be hold both under $A^{(0)\nu}$(non-perturbative EM potential) and 
under $A^{(0)\nu}+\Delta A^{\nu}$, therefore, 
\[
\left. \delta \h{\psi}_{\alpha}^{\dagger}(x^{\prime}) {\Large \backslash} \delta\c{I}_{\mbox{mat}} 
\right|_{(\h{\psi}_{\alpha},\h{\psi}_{\alpha}^{\dagger},A^{\nu})=
(\h{\psi}_{\alpha}[A^{(0)\nu}+\Delta A^{\nu}],\h{\psi}_{\alpha}^{\dagger}[A^{(0)\nu}+\Delta A^{\nu}],A^{(0)\nu}+\Delta A^{\nu})} 
=0,
\]
Taylor expansion leads to:
\[
\sum_{n=0}^{\infty}\frac{1}{n!}
\int \dn{4}{x}_{n}\cdots \int \dn{4}{x}_{1}
\left. \frac{\delta^{n} \left( \delta \h{\psi}_{\alpha}^{\dagger}(x^{\prime}) {\Large \backslash} \delta\c{I}_{\mbox{mat}} \right)  }{\delta A^{\mu_{n}}(x_{n}) \cdots \delta A^{\mu_{1}}(x_{1})} 
\right|_{(\h{\psi}_{\alpha},\h{\psi}_{\alpha}^{\dagger},A^{\nu})=
(\h{\psi}_{\alpha}[A^{(0)\nu}],\h{\psi}_{\alpha}^{\dagger}[A^{(0)\nu}],A^{(0)\nu})} 
\hspace*{-10.0em}
\Delta A^{\mu_{1}}(x_{1}) \cdots  \Delta A^{\mu_{n}}(x_{n}) 
=0,
\]
Considering this equation as the identity with respect to $\Delta A^{\mu}(x)$ results in
\eq(nFunctionalDerivativeA01).
\Eq(nFunctionalDerivativeA02) is proved in the same manner, starting from
\eq(psiDagger_opt).
%
%
\section{Calculation details in \S\ref{sec:oneElectron}}\;
\label{sec:methods}
%
Here we provide the calculation details in \S\ref{sec:oneElectron}, 
including the derivation of the unfamiliar relationship \eq(tradeoff) between 
two types of dipole transition matrix elements.
\subsection{Derivation of the constitutive equations, \eqs(rho_ind_av)-(j_ind_av),
and the susceptibilities, \eq(suscept4rho01), \eqs(suscept4j0)-(suscept4j1)}
\label{cd:1}
%
The incident SP and VP,
$\Delta\phi(\v{r},t)$ and $\Delta A_{i}(\v{r},t)$, are assumed to be 
monochromatic with the angular momentum $\omega$, 
and are expressed using the Coulomb gauge and LWA as follows:
\begin{align}
\label{El}
\Delta\phi(\v{r},t)&=\Delta\phi(\v{r})\cos\omega t 
= \left( \Delta\phi(\v{0}) -  \Delta\v{E}^{(\ell)}(\v{0})\cdot\v{r} \right) \cos\omega t\,,
\\
\label{Et}
\Delta \v{A}(\v{r},t)&=\Delta\v{A}(\v{r})\sin(\omega t+\xi) = -\frac{1}{\omega} \Delta\v{E}^{(t)}(\v{0}) 
\sin(\omega t+\xi)\,,
\end{align}
where $\xi$ is the phase difference between the two incident potentials.
In the spinless one-electron system, 
the linear response theory with \eqs(jind)-(sus01) leads to 
the Heisenberg operators of the induced charge and current densities, 
as follows in the three-element representation: 
\begin{align}
\nonumber
\Delta\h{\rho}(\v{r},t)
=& \int_{-\infty}^{t} \!\!\!\d{t_{1}} \int \dn{3}{r_{1}} 
\left\{
\frac{1}{i\hbar} \left[ \h{\rho}^{(in0)}(\v{r},t)\,,\, \h{\rho}^{(in0)}(\v{r}_{1},t_{1}) \right]  \Delta\phi(\v{r}_{1},t_{1})
\right.
\\
\label{rho_ind}
&\left.
\hspace{0.15\textwidth}
-\frac{1}{i\hbar} \left[ \h{\rho}^{(in0)}(\v{r},t)\,,\, \h{j}^{(in0)}_{i_{1}}(\v{r}_{1},t_{1}) \right]  \Delta A_{i_{1}}(\v{r}_{1},t_{1})
\right\}\,,
\\
\nonumber
\Delta\h{j}_{i}(\v{r},t)
=&
\int_{-\infty}^{t} \!\!\!\d{t_{1}} \int \dn{3}{r_{1}} 
\left\{
\frac{1}{i\hbar} \left[ \h{j}^{(in0)}_{i}(\v{r},t)\,,\, \h{\rho}^{(in0)}(\v{r}_{1},t_{1}) \right]  \Delta\phi(\v{r}_{1},t_{1})
\right.
\\
\label{j_ind}
&\left.
\hspace{0.15\textwidth}
-\frac{1}{i\hbar} \left[ \h{j}^{(in0)}_{i}(\v{r},t)\,,\, \h{j}^{(in0)}_{i_{1}}(\v{r}_{1},t_{1}) \right]  \Delta A_{i_{1}}(\v{r}_{1},t_{1})
\right\}
 -\frac{q}{m} \h{\rho}^{(in0)}(\v{r},t) \Delta A_{i}(\v{r},t) \,.
\end{align}
The last term in \eq(j_ind) originates from {\it the non-relativistic nature} of the system
and is needed to maintain charge conservation law.

Evaluating the expectation value 
of \eqs(rho_ind)-(j_ind)
using the ground state [$\varphi_{0}(\v{r})$ in \eq(states)] and
substituting \eqs(El)-(Et) leads to 
\eqs(rho_ind_av)-(j_ind_av), 
in which the causes of the responses are the two types of electric fields
and their temporal derivatives, defined as
\begin{align}
\label{delE}
&\Delta E^{(\ell)}_{j}(\v{0},t) \equiv \;\;\,\Delta E^{(\ell)}_{j}(\v{0}) \cos\omega t\,,\quad
\Delta E^{(t)}_{j}(\v{0},t) \equiv \;\;\,\Delta E^{(t)}_{j}(\v{0}) \cos(\omega t +\xi)\,,
\\
\label{delEd}
&\Delta \dot{E}^{(\ell)}_{j}(\v{0},t) \equiv \frac{\partial}{\partial t}\Delta E^{(\ell)}_{j}(\v{0},t)\,,
\quad\quad\;\,
\Delta \dot{E}^{(t)}_{j}(\v{0},t) \equiv 
\frac{\partial}{\partial t}\Delta E^{(t)}_{j}(\v{0},t)\,.
\end{align}
In the above, no magnetic response appears because it is
the higher order in the LWA as revealed by Cho \cite{Cho01,Cho02}. 
%
%
To obtain susceptibilities,  \eq(suscept4rho01),\eqs(suscept4j0)-(suscept4j1)
using the two-level model, 
we take the expectation values of \eqs(rho_ind)-(j_ind) using the ground state
$\varphi_{0}(\v{r})$,  
insert the projection operator [the left side of the second equation in \eq(noc)],
between the two operators in the commutators, 
and integrate over the domains of $t_{1}$ and $\v{r}_{1}$.
We assume that the two orbitals are real functions, and form the normalized orthogonal complete set:
\begin{align}
\label{noc}
\int\dn{3}{r} \varphi_{m}(\v{r})\varphi_{n}(\v{r}) = \delta_{m\,n}\,,\quad\quad
\sum_{m} \varphi_{m}(\v{r})\varphi_{m}(\v{r}^{\prime}) = \delta^{3}(\v{r}-\v{r}^{\prime})\,,
\end{align} 
where $\varphi_{m}(\v{r})$ satisfies,  
\begin{align}
\label{states}
\h{H}^{(0)} \varphi_{m}(\v{r}) = \hbar\omega_{m}\,\varphi_{m}(\v{r})\,,\quad (m=0,1)\,.
\end{align}
Having real orbitals infers even temporal parity, such that
there is a null VP (or magnetic field) in the non-perturbed system.
To derive the susceptibilities associated 
with the transversal electric field in \eqs(suscept4rho01)-(suscept4j1),
we use the well-known linear relationship between
the two types of dipole transition matrix elements,
\begin{align}
\label{wellknown}
 \c{C}_{i}&\equiv 
\int \dn{3}{r}\left( 
\partial_{i} \varphi_{1}(\v{r}) \varphi_{0}(\v{r}) - \varphi_{1}(\v{r}) \partial_{i} \varphi_{0}(\v{r}) 
\right)
= 
\frac{2m}{\hbar^{2}}\hbar\Delta\omega_{1}\, \c{D}_{i}\,.
\end{align}
\Eq(wellknown) is derived from the matrix element of Heisenberg equation for dipole charge density:
\begin{align}
\frac{\partial}{\partial t}\, r_{j} \h{\rho}^{(in0)}(\v{r},t)
= 
\frac{1}{i\hbar} 
\left[
 r_{j} \h{\rho}^{(in0)}(\v{r},t)\,,\,\h{H}^{(0)}
\right] \,,
\end{align}
using $\displaystyle \h{\rho}^{(in0)}(\v{r},t) = e^{-\frac{\h{H}^{(0)}t}{i\hbar}} \h{\rho}^{(in0)}(\v{r},0)e^{+\frac{\h{H}^{(0)}t}{i\hbar}}$ and the projection operator, i.e., the second equation in \eq(noc)
satisfying \eq(states).
%
\subsection{ 
Derivation of the spatial average of the susceptibilities, \eqs(suscept4rho_far)-(suscept4j_far)}
\label{cd:3}
%
The following replacements in
\eq(suscept4rho01), \eqs(suscept4j0)-(suscept4j1)
lead to \eqs(suscept4rho_far)-(suscept4j_far):
\begin{align}
\varphi_{0}(\v{r})\varphi_{1}(\v{r})
&\quad\longrightarrow\quad 
\frac{1}{\c{V}}\int\dn{3}{r} \varphi_{0}(\v{r})\varphi_{1}(\v{r}) = 0\,,
\\
\partial_{i} \varphi_{1}(\v{r}) \varphi_{0}(\v{r}) - \varphi_{1}(\v{r}) \partial_{i} \varphi_{0}(\v{r})
&\quad\longrightarrow\quad
\frac{1}{\c{V}}\int\dn{3}{r} \partial_{i} \varphi_{1}(\v{r}) \varphi_{0}(\v{r}) - \varphi_{1}(\v{r}) \partial_{i} \varphi_{0}(\v{r})
=\frac{1}{\c{V}} \c{C}_{i}\,,
\\
\varphi_{0}(\v{r})\varphi_{0}(\v{r})
&\quad\longrightarrow\quad 
\frac{1}{\c{V}}\int\dn{3}{r} \varphi_{0}(\v{r})\varphi_{0}(\v{r}) =\frac{1}{\c{V}}\,.
\end{align}
To derive \eq(suscept4j_far), we additionally use the trade-off relationship
between the two types of dipole transition matrix elements,
\begin{align}
\label{tradeoff}
\c{D}_{i}\, \c{C}_{j} = \delta_{i\,j}.
\end{align}
This is effective in the two-level system with well-defined parity
and derived from the quantum-mechanical commutation relationship: 
\beq
\label{z-pz}
[ r_{i}\,,\, \frac{\hbar}{i}\partial_{j}  ] = i\hbar\,\delta_{ij}\,,\quad\text{i.e.,} \quad
r_{i} \left(   \frac{\hbar}{i}\partial_{j} \cdots \right) +  \frac{\hbar}{-i}\partial_{j} \left(r_{i} \cdots \right) = i\hbar \delta_{ij}\cdots \,. 
\eeq 
Inserting the projection operator 
between $r_{i}$ and $\frac{\hbar}{i}\partial_{j}$,
and eliminating the null integrals caused by mismatched parity result in \eq(tradeoff).
From \eq(wellknown) and \eq(tradeoff), $\c{D}_{i}$ and $\c{C}_{i}$ are specified as
\begin{align}
\label{CDspecify}
\c{D}_{i} = \frac{1}{\c{C}_{i}} = \frac{\hbar}{\sqrt{2m\,\hbar\Delta\omega_{1}}}\,.
\end{align}
(We do not use \eq(CDspecify) in this paper.)
%
%

%

\begin{thebibliography}{00}
%
\bibitem{SpringerFallacy}
I. Banno, 
in 
{\it Progress in Nanophotonics vol. 5} \;edited by T. Yatsui (Springer International Publishing, 2018) Chap.  6.
%
\bibitem{London} 
F. London, {\it Superfluids vol.1, Macroscopic Theory of Superconductivity}
(Dover Publications, Inc., New York, 1950).
%
\bibitem{ABeffect} 
Y. Aharanov and D. Bohm, Phys. Rev. {\bf 115}, 485 (1959).
%
\bibitem{Cho01} K. Cho, J. Phys. Condens. Matter {\bf 20}, 175202 (2008).
%
\bibitem{Cho02} K. Cho, {\it Reconstruction of Macroscopic Maxwell Equations}
 (Springer-Verlag, Berlin, Heidelberg, 2010).
%
\bibitem{DBF}
F. I. Fedorov, Optics and Spectroscopy {\bf 6}, 49 (1959); {\it ibid.} {\bf 6}, 237(1959).
[translated from Russian journal "Optika i Spectroskopiia"];\, 
Ref.\cite{Cho02}, \S{3.4}. 
%
\bibitem{Toyozawa} Y. Toyozawa, {\it The Physics of Elementary Excitations} 
edited by S. Nakajima, Y. Toyozawa, and R. Abe
(Springer-Verlag, Berlin, Heidelberg, 1980). Chap. 2. [This book was translated from
Japanese book "Bussei II" in Iwanami Series of Fundamental Physics (Iwanami Shoten,1978).]
%
\bibitem{Cho00} K. Cho, {\it Optical Response of Nanostructures} (Springer-Verlag, Berlin, Heidelberg, 2003).
%
%
\bibitem{Keller01}
O. Keller, {\it  Quantum Theory of Near-Field Electrodynamics} (Springer, Heidelberg, Dordrecht, London, New York, 2011) Chap. 10.
%
\bibitem{Grassmann}
C. Itzykson and J.-B. Zuber, {\it Quantum Field Theory (International Edition)} 
(MacGraw-Hill Book Co., 1985). Chap. 9.  
; T. Kugo, {\it G\={e}gi-ba no Riron (Quantum Theory of Gauge Fields)} (in Japanese)  (Baifukan, Tokyo, 1989) Chap. 1.
%
\bibitem{EDM2} 
The Mathematical Society of Japan and K. Ito eds.
{\it Encyclopedic Dictionary of Mathematics 2nd Ed.} (MIT Press, 1993).
[This book was translated from Japanese Dictionary 
"Iwanami S\={u}gaku Jiten 3rd Ed." (Iwanami Shoten, 1985).]
\bibitem{Nishijima}
K. Nishijima, {\it Fields and Particles} (W. A. Benjamin, Inc., 1969). Chap. 4.
\bibitem{DFT}
P. Hohenberg and W. Kohn, Phys. Rev. {\bf 136}, 3864 (1964).
\bibitem{Kohn-Sham}
W. Kohn and L. J. Sham, Phys Rev. {\bf 140}, A1133 (1965).
\bibitem{Levy}
M. Levy, Proc. Natl. Acad. Sci. USA {\bf 76}, 6062 (1979).
%
\bibitem{FDTD} 
K. Yee, 
{\it IEEE Transactions on Antennas and Propagation} {\bf14,}  302
(1966).
%
\bibitem{SiLED} T. Kawazoe, M. A. Mueed,  and M. Ohtsu, 
Appl. Phys. B  {\bf 104}, 747
(2011).
%
\bibitem{OhtsuSiLED} M. Ohtsu,
{\it Silicon Light-Emitting Diodes and Lasers} 
(Springer International Publishing, Switsland, 2016).
%
\bibitem{SiLaser} T. Kawazoe, M. Ohtsu, K. Akahane, and N. Yamamoto, 
{Appl. Phys. B}  {\bf 107}, 659
(2012).
%
\bibitem{NFCVD} T. Kawazoe, Y. Yamamoto, and M. Ohtsu, 
{Appl. Phys. Lett.} {\bf 79}, 1184
(2001).
%
\bibitem{NFLitho} H. Yonemitsu, T. Kawazoe, K. Kobayashi, and M. Ohtsu, 
{J. Photolumin.} {\bf 122}, 230
(2007).
%
%
\bibitem{NFEtching} T. Yatsui, K. Hirata, W. Nomura, Y. Tabata, and M. Ohtsu, 
{Appl. Phys. B}  {\bf 93}, 55
(2008).
%
\bibitem{NFUpConv} T. Kawazoe, H. Fujiwara, K. Kobayashi, and M. Ohtsu,  
{IEEE J. of Selected Topics in Quantum Electronics} {\bf 15}, 1380
(2009).
%
\bibitem{PlsShpMs}
H. Fujiwara, T. Kawazoe, and M. Ohtsu,
{Appl. Phys. B} {\bf 100}, 85
(2010).
%
\bibitem{nanoFountain}
T. Kawazoe, K. Kobayashi, and M. Ohtsu,
{Appl. Phys. Lett.} {\bf 86}, 103102-1
(2005).
%
\bibitem{nanoPhotonicDevice}
T. Kawazoe, M. Ohtsu, S. Aso, Y. Sawado, Y. Hosoda, K. Yoshizawa, K. Akahane, N. Yamamoto, and M. Naruse, 
{Appl. Phys. B} {\bf103}, 537
(2011). 
%
\bibitem{MO}
N. Tate, T. Kawazoe, W. Nomura, and M. Ohtsu,
{Scientific Reports} {\bf 5}, 12762-1
(2015).
%
%
\end{thebibliography}
\end{document}


\preprint{APS/123-QED}
%
\title{
{\small
Supplementary Document for\\
"Theory of Single Susceptibility for Near-field Optics\\
Equally Associated with Scalar and Vector Potentials"\\[\baselineskip]}
Check for Charge Conservation Law and Gauge Invariance
}
\author{Itsuki Banno}
\email{banno@yamanashi.ac.jp}
\affiliation{
Interdisciplinary Graduate School of Medicine and Engineering,
University of Yamanashi,
4-3-11 Takeda, Kofu, Yamanashi 400-8511, Japan }
\date{\today}
\begin{abstract}
Charge conservation law and gauge invariance are explicitly checked
for the linear and nonlinear single susceptibilities derived in the main text.
\end{abstract}
\pacs
{
78.67.-n, 
78.20.Bh, 
41.20.-q, 
42.25.Ja 
}
\keywords{
single susceptibility,
non-resonant effect, 
optical near field, 
response function,
electromagnetic potential
}
\maketitle
%
%
%
\renewcommand{\thesection}{\Alph{section}}
\renewcommand{\thesubsection}{\Alph{section} \arabic{subsection}}
 \renewcommand{\theequation}{%
    \thesection \arabic{equation}}
\setcounter{section}{18}
\section{Charge Conservation Law and Gauge Invariance of Linear and Nonlinear Four-Element Single Susceptibility}
\label{sec:ccl_gi_explicite}
%
\subsection{Linear single susceptibility: Eq.(40) in the main text}
%
To show that four-element linear single susceptibility guarantees the charge conservation law, Eq.(24) in the main text,
suppose the four-element divergence of Eq.(40) in the main text, considering  $\partial_{\mu}\h{j}^{(in0)\,\mu}(x)=0$,
\beqna
\label{sus01_ccl}
\partial_{\mu}
\h{\chi}^{\mu}_{\;\;\mu_{1}}(x,x_{1}) 
&=&  
\frac{-q}{mc^{2}}\, \delta(ct-ct_{1})
\t{\delta}^{\mu}_{\;\;\mu_{1}} \partial_{\mu}
\left(
\delta^{3}(x-x_{1})\, \h{j}^{(in0)\,0}(x)
\right)
\\
\nonumber
&&
+\frac{1}{i\hbar c^{2} } \delta(ct-ct_{1}) \left[  \h{j}^{(in0)\,0}(x), \h{j}^{(in0)}_{\quad\mu_{1}}(x_{1}) \right]
\\
\nonumber
&=&0.
\eeqna
In the second term of the second hand, we use the following commutation relationship at the same time
: 
\beqna
\label{j0jmu}
\delta(ct-ct_{1})
\left[  \h{j}^{(in0)\,0}(x), \h{j}^{(in0)}_{\quad\mu_{1}}(x_{1}) \right]
&=& 
- ihc^{2} \frac{-q}{mc^{2}}
\delta(ct-ct_{1})
\t{\delta}^{\mu}_{\mu_{1}} \partial_{\mu}
\left(  \delta^{3}(x-x_{1})   \h{j}^{(in0)\,0}(x)\right)
\\
\nonumber
&=&
- ihc^{2} \frac{-q}{mc^{2}}
\delta(ct-ct_{1})
\t{\delta}^{\mu}_{\mu_{1}} 
\left(  \partial_{\mu}
\delta^{3}(x-x_{1})  \right) \h{j}^{(in0)\,0}(x_{1})
\eeqna
%
%
The proof of \eq(j0jmu) is as follows:
If $\mu_{1}=0$ in the left hand side of \Eq(j0jmu),
it is the commutator between charge density operator at the same time, and
is zero. 
\beq
\label{j0jmu_0}
\left[  \h{j}^{(in0)\,0}(x), \h{j}^{(in0)}_{\quad 0}(x_{1}) \right]_{t_{1}=t}
=c^{2}q^{2}\left[  \h{\psi}_{\alpha}^{\dagger}(x)\h{\psi}_{\alpha}(x)\,,\,   \h{\psi}_{\alpha}^{\dagger}(x_{1})\h{\psi}_{\alpha}(x_{1}) \right]_{t_{1}=t}
=0.
\eeq
One may check \eq(j0jmu_0) by a straightforward calculation using the anti-commutation relation of
electron field operators at the same time,
\[
\left[ \h{\psi}_{\alpha}(x)\,,\,\h{\psi}_{\alpha}^{\dagger}(x_{1}) \right]_{+,\,t_{1}=t} = \delta^{3}(x-x_{1}).
\]

Next, if $\mu_{1}=i \in\{1,2,3\}$ in the left hand side of \Eq(j0jmu),
the commutator in three-element representation becomes as follows:
\beqna
\nonumber
&&\left[  \h{j}^{(in0)\,0}(x), \h{j}^{(in0)}_{\quad \mu_{1}=i}(x_{1}) \right]_{t_{1}=t}
\\
\nonumber
&=&-cq \frac{q}{2m}
\left[  \h{\psi}_{\alpha}^{\dagger}(x)\h{\psi}_{\alpha}(x)\,,\,   
\h{\psi}_{\alpha}^{\dagger}(x_{1})\left( \frac{\hbar}{i}\partial_{i}^{1}-qA^{(0)}_{i}(x_{1})  \right)  \h{\psi}_{\alpha}(x_{1})
\right.
\\
&&\hspace*{0.28\textwidth}
\left.
+
\left( \left( \frac{\hbar}{-i}\partial_{i}^{1}-qA^{(0)}_{i}(x_{1})  \right) \h{\psi}_{\alpha}^{\dagger}(x_{1})\right)  \h{\psi}_{\alpha}(x_{1})
 \right]_{t_{1}=t}
\eeqna
As the term includes $qA^{(0)}_{i}(x_{1}) $ is zero following \eq(j0jmu_0),
let us treat the term including the derivative. 
\beqna
\nonumber
&&\left[  \h{j}^{(in0)\,0}(x), \h{j}^{(in0)}_{\quad \mu_{1}=i}(x_{1}) \right]_{t_{1}=t}
\\
\nonumber
&=&
-\frac{cq}{2} i\hbar c^{2}\frac{-q}{mc^{2}}
\lim_{x_{1}^{\bullet}\to x_{1}} \partial_{i}^{1\,\bullet}
\left[
\h{\psi}_{\alpha}^{\dagger}(x)\h{\psi}_{\alpha}(x)\,,\, 
\h{\psi}_{\alpha}^{\dagger}(x_{1}) \h{\psi}_{\alpha}(x_{1}^{\bullet})
- \h{\psi}_{\alpha}^{\dagger}(x_{1}^{\bullet}) \h{\psi}_{\alpha}(x_{1})
\right]_{t_{1}=t}
\\
\nonumber
&=&
-\frac{cq}{2} i\hbar c^{2}\frac{-q}{mc^{2}}
\\
\nonumber
&&\quad
\lim_{x_{1}^{\bullet}\to x_{1}} \partial_{i}^{1\,\bullet}
\left(
\delta^{3}(x-x_{1}) 
\left(  \h{\psi}_{\alpha}^{\dagger}(x) \h{\psi}_{\alpha}(x_{1}^{\bullet}) + \h{\psi}_{\alpha}^{\dagger}(x_{1}^{\bullet}) \h{\psi}_{\alpha}(x)   
\right)
\right.
\\
\nonumber
&&\hspace*{0.13\textwidth}
\left. 
-\delta^{3}(x-x_{1}^{\bullet}) 
\left(  \h{\psi}_{\alpha}^{\dagger}(x) \h{\psi}_{\alpha}(x_{1}) + \h{\psi}_{\alpha}^{\dagger}(x_{1}) \h{\psi}_{\alpha}(x)    
\right) 
\right)_{t_{1}=t} 
\\
\nonumber
&=&
-\frac{cq}{2} i\hbar c^{2}\frac{-q}{mc^{2}}
\\
\nonumber
&&\quad
\delta^{3}(x-x_{1}) 
\partial_{i}\left(  \h{\psi}_{\alpha}^{\dagger}(x) \h{\psi}_{\alpha}(x_{1}) + \h{\psi}_{\alpha}^{\dagger}(x_{1}) \h{\psi}_{\alpha}(x) \right)
\\
\nonumber
&&\hspace*{0.25\textwidth}
+ \left( \partial_{i}\delta^{3}(x-x_{1}) \right)
\left(  \h{\psi}_{\alpha}^{\dagger}(x) \h{\psi}_{\alpha}(x_{1}) + \h{\psi}_{\alpha}^{\dagger}(x_{1}) \h{\psi}_{\alpha}(x)    
\right)_{t_{1}=t}
\\
\nonumber
&=&
-\frac{cq}{2} i\hbar c^{2}\frac{-q}{mc^{2}}
\partial_{i}
\left( \delta^{3}(x-x_{1}) 
\left(  \h{\psi}_{\alpha}^{\dagger}(x) \h{\psi}_{\alpha}(x_{1}) + \h{\psi}_{\alpha}^{\dagger}(x_{1}) \h{\psi}_{\alpha}(x) \right)  
\right)_{t_{1}=t}
\\
\label{j0jmu_1p}
&=&
-cq i\hbar c^{2}\frac{-q}{mc^{2}}
\partial_{i}
\left( \delta^{3}(x-x_{1})  \h{\psi}_{\alpha}^{\dagger}(x) \h{\psi}_{\alpha}(x)   
\right)
=
- i\hbar c^{2}\frac{-q}{mc^{2}}
\t{\delta}_{\mu_{1}}^{\;\mu} \partial_{\mu}
\left( \delta^{3}(x-x_{1})  \h{j}^{(in0)\,0}(x) \right)
\\
\nonumber
&=&
-cq i\hbar c^{2}\frac{-q}{mc^{2}}
\partial_{i}
\left( \delta^{3}(x-x_{1})  \h{\psi}_{\alpha}^{\dagger}(x_{1}) \h{\psi}_{\alpha}(x_{1})
\right)_{t_{1}=t}
\!\!\!\!\!\!\!\!=
- i\hbar c^{2}\frac{-q}{mc^{2}}
\t{\delta}_{\mu_{1}}^{\;\mu} 
\left( \partial_{\mu}\delta^{3}(x-x_{1}) \right)  \h{j}^{(in0)\,0}(x_{1})_{t_{1}=t}\, ,
\\
\label{j0jmu_1}
&&
\eeqna
where the last two-way expressions \eqs(j0jmu_1p)-(j0jmu_1) are 
in four-element representation instead of three-element representation.
Summarizing \eqs(j0jmu_0)-(j0jmu_1) result in \eq(j0jmu).
As a result, the present four-element linear susceptibility,Eq.(40) in the main text maintains the charge conservation law, Eq.(24) in the main text.

For the proof for the gauge invariance, Eq.(25) in the main text, of the linear susceptibility,  
suppose the four-element divergence with respect to $x_{1}$.
Then, using  \eq(j0jmu) with the replacement,
$x \leftrightarrow x_{1}$ and the relation, 
$\t{\delta}^{\mu_{1}}_{\mu}\partial_{\mu_{1}}= -\t{\delta}_{\mu_{1}}^{\mu}\partial^{\mu_{1}}$,
one may obtain:
\beqna
\label{sus01_gi}
\partial^{\mu_{1}}
\h{\chi}^{\mu}_{\;\;\mu_{1}}(x,x_{1}) 
&=&  
\frac{-q}{mc^{2}}\, \delta(ct-ct_{1})
\left( 
\t{\delta}^{\mu}_{\;\;\mu_{1}} 
\partial^{\mu_{1}}
\delta^{3}(x-x_{1}) \right)
\, \h{j}^{(in0)\,0}(x)
\\
\nonumber
&&
-\frac{1}{i\hbar c^{2} } \delta(ct-ct_{1}) \left[  \h{j}^{(in0)\,\mu}(x), \h{j}^{(in0)}_{\quad0}(x_{1}) \right]
\\
\nonumber
&=&0.
\eeqna
As shown above, the linear susceptibility Eq.(40) in the main text maintains the gauge invariance Eq.(25) in the main text.
%
\subsection{Second order nonlinear single susceptibility: Eq.(41) in the main text}
%
Next, let us show that the charge conservation law is satisfied by the
second order nonlinear single susceptibility,  Eq.(41) in the main text.
Operating 
$\partial_{\mu} = \delta_{\mu}^{0}\partial_{0}+ \delta_{\mu}^{1}\partial_{1}
+ \delta_{\mu}^{2}\partial_{2}+ \delta_{\mu}^{3}\partial_{3} $
to  Eq.(41) in the main text and considering $\partial_{\mu} \h{j}^{(in0)\,\mu}(x)=0$
(the charge conservation law for
the current density operator in the
non-interacting system), 
the surviving terms are those the operator $\partial_{\mu}$ operates on
the step function or delta function in front of the commutator, 
and operates on $\h{j}^{(in0)\,0}(x)$ in the commutator.
\begin{subequations}
\beqna
\nonumber
&& 
\hspace*{-8.0em}
\partial_{\mu}\:
2!\,
\h{\chi}^{\mu}_{\;\;\mu_{1}\mu_{2}}(x,x_{1},x_{2})
=
\\
\label{sus02_01}
\frac{1}{i\hbar c ^{2}}\, \frac{-q}{mc^{2}}
&&
\left\{ 
\delta(ct-ct_{1}) \theta(ct-ct_{2})\quad
\t{\delta}^{\mu}_{\;\;\mu_{1}} 
\partial_{\mu}\left(
\delta^{3}(x-x_{1}) 
\left[ \h{j}^{(in0)\,0}(x), \h{j}^{(in0)}_{\quad\mu_{2}}(x_{2}) \right] 
\right)
\right.
\\
\label{sus02_02}
&&+ 
\delta(ct-ct_{2}) \theta(ct-ct_{1})\quad
\t{\delta}^{\mu}_{\;\;\mu_{2}} 
\partial_{\mu}\left(
\delta^{3}(x-x_{2}) 
\left[ \h{j}^{(in0)\,0}(x), \h{j}^{(in0)}_{\quad\mu_{1}}(x_{1}) \right] 
\right)
\\
\label{sus02_03}
&&+
\left.
\delta(ct-ct_{1})\delta(ct_{1}-ct_{2})
\t{\delta}_{\mu_{1}\,\mu_{2}} \quad\:\delta^{3}(x_{1}-x_{2}) 
\left[ \h{j}^{(in0)\,0}(x), \h{j}^{(in0)}_{\quad 0}(x_{1}) \right] 
\right\}
\\
\label{sus02_04}
+\left(\frac{1}{i\hbar c^{2} }\right)^{2} 
&&
\left\{
\delta(ct-ct_{1}) \theta(ct_{1}-ct_{2})
\left[ \left[ \h{j}^{(in0)\,0}(x), \h{j}^{(in0)}_{\quad \mu_{1}}(x_{1}) \right], 
\h{j}^{(in0)}_{\quad \mu_{2}}(x_{2}) \right]   
\right.
\\
\label{sus02_05}
&&\!\!+ 
\left.
\delta(ct-ct_{2}) \theta(ct_{2}-ct_{1})
\left[ \left[ \h{j}^{(in0)\,0}(x), \h{j}^{(in0)}_{\quad \mu_{2}}(x_{2}) \right], 
\h{j}^{(in0)}_{\quad \mu_{1}}(x_{1}) \right]   
\right\} .
\eeqna
\end{subequations}
Applying \eq(j0jmu), the third term (\ref{sus02_03}) vanishes, and
the fourth and fifth terms (\ref{sus02_04}) and (\ref{sus02_05})
cancel the first and second terms (\ref{sus02_01}) and (\ref{sus02_02}), respectively.

As a result, the second order nonlinear single susceptibility operator Eq.(41) in the main text
maintains the charge conservation law, Eq.(24) in the main text.

To check the gauge invariance of the second order nonlinear single susceptibility operator,
let us operate $\partial^{\mu_{1}}$ to Eq.(41) in the main text.
\begin{subequations}
\beqna
\nonumber
&& 
\hspace*{-5.0em}
\partial^{\mu_{1}}\:
2!\,
\h{\chi}^{\mu}_{\;\;\mu_{1}\mu_{2}}(x,x_{1},x_{2})
=
\\
\label{sus02g_01}
\frac{1}{i\hbar c ^{2}}\, \frac{-q}{mc^{2}}
&&
\left\{ 
\delta(ct-ct_{1}) \theta(ct-ct_{2})\;\;\:
\left(\t{\delta}^{\mu}_{\;\;\mu_{1}}  \partial^{\mu_{1}}
\delta^{3}(x-x_{1}) \right)
\left[ \h{j}^{(in0)\,0}(x), \h{j}^{(in0)}_{\quad\mu_{2}}(x_{2}) \right] 
\right.
\\
\label{sus02g_02}
&&- 
\delta(ct-ct_{2}) \delta(ct-ct_{1})\quad
\t{\delta}^{\mu}_{\;\;\mu_{2}} \quad \delta^{3}(x-x_{2}) 
\quad\;\:
\left[ \h{j}^{(in0)\,0}(x), \h{j}^{(in0)}_{\quad 0}(x_{1}) \right] 
\\
\label{sus02g_03}
&&\!+
\left.
\theta(ct-ct_{1})\delta(ct_{1}-ct_{2})
\t{\delta}_{\mu_{1}\,\mu_{2}} 
\partial^{\mu_{1}}\left(
\delta^{3}(x_{1}-x_{2}) 
\left[ \h{j}^{(in0)\,\mu}(x), \h{j}^{(in0)}_{\quad 0}(x_{1}) \right] 
\right)
\right\}
\\
\nonumber
+\left(\frac{1}{i\hbar c^{2} }\right)^{2} 
&&
\left\{
\left(-\delta(ct-ct_{1}) \theta(ct_{1}-ct_{2})+\theta(ct-ct_{1}) \delta(ct_{1}-ct_{2})\right)
\right.
\\
&&\hspace{10.4em}
\label{sus02g_04}
\left[ \left[ \h{j}^{(in0)\,\mu}(x), \h{j}^{(in0)}_{\quad 0}(x_{1}) \right], 
\h{j}^{(in0)}_{\quad \mu_{2}}(x_{2}) \right]   
\\
\label{sus02g_05}
&&\!\!-
\left.
\theta(ct-ct_{2}) \delta(ct_{2}-ct_{1})
\left[ \left[ \h{j}^{(in0)\,\mu}(x), \h{j}^{(in0)}_{\quad \mu_{2}}(x_{2}) \right], 
\h{j}^{(in0)}_{\quad 0}(x_{1}) \right]   
\right\} .
\eeqna
\end{subequations}
Replacing the fifth term (\ref{sus02g_05}) using the next Jacobi identity:
\beqna
\label{jacobi05}
&&
\left[ \left[ \h{j}^{(in0)\,\mu}(x), \h{j}^{(in0)}_{\quad \mu_{2}}(x_{2}) \right], 
\h{j}^{(in0)}_{\quad 0}(x_{1}) \right]   
\\
\nonumber
&=&
-\left[ \left[ \h{j}^{(in0)}_{\quad \mu_{2}}(x_{2}),\h{j}^{(in0)}_{\quad 0}(x_{1}) \right], 
\h{j}^{(in0)\,\mu}(x) \right]      
-\left[ \left[ \h{j}^{(in0)}_{\quad 0}(x_{1}),\h{j}^{(in0)\,\mu}(x) \right], 
\h{j}^{(in0)}_{\quad \mu_{2}}(x_{2}) \right],  
\eeqna
then, the first term in the right hand side of \eq(jacobi05) with \eq(j0jmu)
offsets the term (\ref{sus02g_03}), and 
the second term in the right hand side of \eq(jacobi05) offsets the
second term in  (\ref{sus02g_04}).
The first term in (\ref{sus02g_04}) offsets the first term, (\ref{sus02g_01}),
considering the commutation relation at the simultaneous time,
$\delta(ct-ct_{1})\left[ \h{j}^{(in0)\,\mu}(x), \h{j}^{(in0)}_{\quad 0}(x_{1}) \right]$
and \eq(j0jmu)(remark the change of upper or lower subscript).
The second term (\ref{sus02g_02}) vanishes by means of \eq(j0jmu).

As a result, the second order nonlinear single susceptibility operator Eq.(41) in the main text
maintains the gauge invariance, Eq.(25) in the main text.
%
\subsection{Third order nonlinear single susceptibility: Eq.(42) in the main text}
%
With respect to the third order nonlinear single susceptibility, let us check the
charge conservation law.
Operating $\partial_{\mu}$ to Eq.(42) in the main text,
\begin{footnotesize}
\begin{subequations}
\beqna
\nonumber
&& 
\partial_{\mu}\,3!\,
\h{\chi}^{\mu}_{\;\;\mu_{1}\mu_{2}\mu_{3}}(x,x_{1},x_{2},x_{3}) 
=
\\
\nonumber
&&
\frac{1}{i\hbar c^{2} }  \left(\frac{-q}{mc^{2}}\right)^{2} 
\\
\label{sus03_01}
&&\left\{
\theta(ct-ct_{2}) \delta(ct-ct_{1}) \delta(ct_{2}-ct_{3}) 
\t{\delta}_{\mu_{2}\,\mu_{3}} \delta^{3}(x_{2}-x_{3}) 
\t{\delta}^{\mu}_{\;\;\mu_{1}} \partial_{\mu} 
\left(
\delta^{3}(x-x_{1}) 
\left[ \h{j}^{(in0)\,0}(x), \h{j}^{(in0)}_{\quad\;\; 0}(x_{2}) \right] 
\right)
\right.
\\
\label{sus03_02}
&&+
\theta(ct-ct_{3}) \delta(ct-ct_{2}) \delta(ct_{3}-ct_{1}) 
\t{\delta}_{\mu_{3}\,\mu_{1}} \delta^{3}(x_{3}-x_{1}) 
\t{\delta}^{\mu}_{\;\;\mu_{2}} \partial_{\mu} 
\left( \delta^{3}(x-x_{2})
\left[ \h{j}^{(in0)\,0}(x), \h{j}^{(in0)}_{\quad\;\; 0}(x_{3}) \right]  
\right)
\\
\label{sus03_03}
&&+
\left.
\theta(ct-ct_{1}) \delta(ct-ct_{3}) \delta(ct_{1}-ct_{2}) 
\t{\delta}_{\mu_{1}\,\mu_{2}} \delta^{3}(x_{1}-x_{2}) 
\t{\delta}^{\mu}_{\;\;\mu_{3}} \partial_{\mu} 
\left( 
\delta^{3}(x-x_{3})
\left[ \h{j}^{(in0)\,0}(x), \h{j}^{(in0)}_{\quad\;\; 0}(x_{1}) \right]  
\right) 
\right\}
%
%
\\
\nonumber
&&+
\left(\frac{1}{i\hbar c^{2}}\right)^{2} \frac{-q}{mc^{2}}
\\
\label{sus03_04}
&&
\left\{
\delta(ct-ct_{1})
\theta(ct_{1}-ct_{2}) 
\theta(ct_{2}-ct_{3})
\t{\delta}^{\mu}_{\;\;\mu_{1}}
\partial_{\mu}
\left( 
\delta^{3}(x-x_{1})
\left[ \left[ \h{j}^{(in0)\,0}(x), \h{j}^{(in0)}_{\quad \mu_{2}}(x_{2}) \right], 
\h{j}^{(in0)}_{\quad \mu_{3}}(x_{3}) \right]  
\right)
\right.
\\
\label{sus03_05}
&&+
\delta(ct-ct_{1})
\theta(ct_{1}-ct_{3}) 
\theta(ct_{3}-ct_{2})
\t{\delta}^{\mu}_{\;\;\mu_{1}}
\partial_{\mu}
\left(
\delta^{3}(x-x_{1})
\left[ \left[ \h{j}^{(in0)\,0}(x), \h{j}^{(in0)}_{\quad \mu_{3}}(x_{3}) \right], 
\h{j}^{(in0)}_{\quad \mu_{2}}(x_{2}) \right]  
\right)
\\
\label{sus03_06}
&&+
\delta(ct-ct_{2})
\theta(ct_{2}-ct_{3}) 
\theta(ct_{3}-ct_{1})
\t{\delta}^{\mu}_{\;\;\mu_{2}}
\partial_{\mu}
\left(
\delta^{3}(x-x_{2})
\left[ \left[ \h{j}^{(in0)\,0}(x), \h{j}^{(in0)}_{\quad \mu_{3}}(x_{3}) \right], 
\h{j}^{(in0)}_{\quad \mu_{1}}(x_{1}) \right]  
\right)
\\
\label{sus03_07}
&&+
\delta(ct-ct_{2})
\theta(ct_{2}-ct_{1}) 
\theta(ct_{1}-ct_{3})
\t{\delta}^{\mu}_{\;\;\mu_{2}}
\partial_{\mu}
\left(
\delta^{3}(x-x_{2})
\left[ \left[ \h{j}^{(in0)\,0}(x), \h{j}^{(in0)}_{\quad \mu_{1}}(x_{1}) \right], 
\h{j}^{(in0)}_{\quad \mu_{3}}(x_{3}) \right]  
\right)
\\
\label{sus03_08}
&&+
\delta(ct-ct_{3})
\theta(ct_{3}-ct_{1}) 
\theta(ct_{1}-ct_{2})
\t{\delta}^{\mu}_{\;\;\mu_{3}}
\partial_{\mu}
\left(
\delta^{3}(x-x_{3})
\left[ \left[ \h{j}^{(in0)\,0}(x), \h{j}^{(in0)}_{\quad \mu_{1}}(x_{1}) \right], 
\h{j}^{(in0)}_{\quad \mu_{2}}(x_{2}) \right]  
\right)
\\
\label{sus03_09}
&&+
\delta(ct-ct_{3})
\theta(ct_{3}-ct_{2}) 
\theta(ct_{2}-ct_{1})
\t{\delta}^{\mu}_{\;\;\mu_{3}}
\partial_{\mu}
\left(
\delta^{3}(x-x_{3})
\left[ \left[ \h{j}^{(in0)\,0}(x), \h{j}^{(in0)}_{\quad \mu_{2}}(x_{2}) \right], 
\h{j}^{(in0)}_{\quad \mu_{1}}(x_{1}) \right]  
\right)
\\
\label{sus03_10}
&&+
\delta(ct-ct_{1}) \delta(ct_{1}-ct_{2})\theta(ct_{2}-ct_{3})
\t{\delta}_{\mu_{1}\,\mu_{2}}\delta^{3}(x_{1}-x_{2})
\left[ \left[ \h{j}^{(in0)\,0}(x), \h{j}^{(in0)}_{\quad 0}(x_{1}) \right], 
\h{j}^{(in0)}_{\quad \mu_{3}}(x_{3}) \right]  
\\
\label{sus03_11}
&&+
\delta(ct-ct_{2}) \delta(ct_{2}-ct_{3})\theta(ct_{3}-ct_{1})
\t{\delta}_{\mu_{2}\,\mu_{3}}\delta^{3}(x_{2}-x_{3})
\left[ \left[ \h{j}^{(in0)\,0}(x), \h{j}^{(in0)}_{\quad 0}(x_{2}) \right], 
\h{j}^{(in0)}_{\quad \mu_{1}}(x_{1}) \right]  
\\
\label{sus03_12}
&&+
\delta(ct-ct_{3}) \delta(ct_{3}-ct_{1})\theta(ct_{1}-ct_{2})
\t{\delta}_{\mu_{3}\,\mu_{1}}\delta^{3}(x_{3}-x_{1})
\left[ \left[ \h{j}^{(in0)\,0}(x), \h{j}^{(in0)}_{\quad 0}(x_{3}) \right], 
\h{j}^{(in0)}_{\quad \mu_{2}}(x_{2}) \right]  
\\
\label{sus03_13}
&&+
\delta(ct-ct_{1}) \theta(ct_{1}-ct_{2})\delta(ct_{2}-ct_{3})
\t{\delta}_{\mu_{2}\,\mu_{3}}\delta^{3}(x_{2}-x_{3})
\left[ \left[ \h{j}^{(in0)\,0}(x), \h{j}^{(in0)}_{\quad \mu_{1}}(x_{1}) \right], 
\h{j}^{(in0)}_{\quad 0}(x_{2}) \right]
\\
\label{sus03_14}
&&+
\delta(ct-ct_{2}) \theta(ct_{2}-ct_{3})\delta(ct_{3}-ct_{1})
\t{\delta}_{\mu_{3}\,\mu_{1}}\delta^{3}(x_{3}-x_{1})
\left[ \left[ \h{j}^{(in0)\,0}(x), \h{j}^{(in0)}_{\quad \mu_{2}}(x_{2}) \right], 
\h{j}^{(in0)}_{\quad 0}(x_{3}) \right]
\\
\label{sus03_15}
&&\left.+
\delta(ct-ct_{3}) \theta(ct_{3}-ct_{1})\delta(ct_{1}-ct_{2})
\t{\delta}_{\mu_{1}\,\mu_{2}}\delta^{3}(x_{1}-x_{2})
\left[ \left[ \h{j}^{(in0)\,0}(x), \h{j}^{(in0)}_{\quad \mu_{3}}(x_{3}) \right], 
\h{j}^{(in0)}_{\quad 0}(x_{1}) \right]
\right\}
%
%
\\
\nonumber
&&+
\left(\frac{1}{i\hbar c^{2} }\right)^{3}
\\
\label{sus03_16}
&&
\left\{ 
\delta(ct-ct_{1}) \theta(ct_{1}-ct_{2})  \theta(ct_{2}-ct_{3})
\left[ \left[ \left[ \h{j}^{(in0)\,0}(x), \h{j}^{(in0)}_{\quad \mu_{1}}(x_{1}) \right], 
\h{j}^{(in0)}_{\quad \mu_{2}}(x_{2}) \right] ,
\h{j}^{(in0)}_{\quad \mu_{3}}(x_{3}) \right]
\right.
\\
\label{sus03_17}
&&+
\delta(ct-ct_{1}) \theta(ct_{1}-ct_{3})  \theta(ct_{3}-ct_{2})
\left[ \left[ \left[ \h{j}^{(in0)\,0}(x), \h{j}^{(in0)}_{\quad \mu_{1}}(x_{1}) \right], 
\h{j}^{(in0)}_{\quad \mu_{3}}(x_{3}) \right] ,
\h{j}^{(in0)}_{\quad \mu_{2}}(x_{2}) \right]
\\
\label{sus03_18}
&&+
\delta(ct-ct_{2}) \theta(ct_{2}-ct_{3})  \theta(ct_{3}-ct_{1})
\left[ \left[ \left[ \h{j}^{(in0)\,0}(x), \h{j}^{(in0)}_{\quad \mu_{2}}(x_{2}) \right], 
\h{j}^{(in0)}_{\quad \mu_{3}}(x_{3}) \right] ,
\h{j}^{(in0)}_{\quad \mu_{1}}(x_{1}) \right]
\\
\label{sus03_19}
&&+
\delta(ct-ct_{2}) \theta(ct_{2}-ct_{1})  \theta(ct_{1}-ct_{3})
\left[ \left[ \left[ \h{j}^{(in0)\,0}(x), \h{j}^{(in0)}_{\quad \mu_{2}}(x_{2}) \right], 
\h{j}^{(in0)}_{\quad \mu_{1}}(x_{1}) \right] ,
\h{j}^{(in0)}_{\quad \mu_{3}}(x_{3}) \right]
\\
\label{sus03_20}
&&+
\delta(ct-ct_{3}) \theta(ct_{3}-ct_{1})  \theta(ct_{1}-ct_{2})
\left[ \left[ \left[ \h{j}^{(in0)\,0}(x), \h{j}^{(in0)}_{\quad \mu_{3}}(x_{3}) \right], 
\h{j}^{(in0)}_{\quad \mu_{1}}(x_{1}) \right] ,
\h{j}^{(in0)}_{\quad \mu_{2}}(x_{2}) \right]
\\
\label{sus03_21}
&&\left.+
\delta(ct-ct_{3}) \theta(ct_{3}-ct_{2})  \theta(ct_{2}-ct_{1})
\left[ \left[ \left[ \h{j}^{(in0)\,0}(x), \h{j}^{(in0)}_{\quad \mu_{3}}(x_{3}) \right], 
\h{j}^{(in0)}_{\quad \mu_{2}}(x_{2}) \right] ,
\h{j}^{(in0)}_{\quad \mu_{1}}(x_{1}) \right]
\right\}.
\eeqna
\end{subequations}
\end{footnotesize}
The term (\ref{sus03_16}) offsets the term (\ref{sus03_04}), applying \eq(j0jmu) to 
the most inner commutator in (\ref{sus03_16}).
In the same manner, the terms (\ref{sus03_17})-(\ref{sus03_21}),respectively,
offsets the terms (\ref{sus03_05})-(\ref{sus03_09}), applying \eq(j0jmu).
The terms (\ref{sus03_10})-(\ref{sus03_12}) vanishes, applying \eq(j0jmu)
to the inner commutator at the simultaneous time.
The term (\ref{sus03_13}) offsets the term (\ref{sus03_01}),
applying \eq(j0jmu) to the most inner commutator.
In the same manner, the terms (\ref{sus03_14})-(\ref{sus03_15}), respectively,
offsets (\ref{sus03_02})-(\ref{sus03_03}), using \eq(j0jmu).

As a result, the third order nonlinear single susceptibility operator Eq.(42) in the main text
maintains the charge conservation law, Eq.(24) in the main text.

To check the gauge invariance of the third order nonlinear single susceptibility operator,
let us operate $\partial^{\mu_{1}}$ to Eq.(42) in the main text.
\begin{footnotesize}
\begin{subequations}
\beqna
\nonumber
&& 
\partial^{\mu_{1}}\,
3!\,
\h{\chi}^{\mu}_{\;\;\mu_{1}\mu_{2}\mu_{3}}(x,x_{1},x_{2},x_{3}) 
=  
\\
\nonumber
&&
\frac{1}{i\hbar c^{2} }  \left(\frac{-q}{mc^{2}}\right)^{2} 
\\
\label{sus03g_01}
&&\left\{
\theta(ct-ct_{2}) \delta(ct-ct_{1}) \delta(ct_{2}-ct_{3}) 
\left( 
\t{\delta}^{\mu}_{\;\;\mu_{1}} 
\partial^{\mu_{1}}
\delta^{3}(x-x_{1}) \right)
\t{\delta}_{\mu_{2}\,\mu_{3}} \delta^{3}(x_{2}-x_{3}) 
\left[ \h{j}^{(in0)\,0}(x), \h{j}^{(in0)}_{\quad\;\; 0}(x_{2}) \right] 
\right.
\\
\label{sus03g_02}
&&+
\theta(ct-ct_{3}) \delta(ct-ct_{2}) \delta(ct_{3}-ct_{1}) 
\t{\delta}^{\mu}_{\;\;\mu_{2}} \delta^{3}(x-x_{2}) 
\left(
\t{\delta}_{\mu_{3}\,\mu_{1}} 
\partial^{\mu_{1}}
\delta^{3}(x_{3}-x_{1}) 
\right)
\left[ \h{j}^{(in0)\,0}(x), \h{j}^{(in0)}_{\quad\;\; 0}(x_{3}) \right] 
\\
\label{sus03g_03}
&&+
\left.
\theta(ct-ct_{1}) \delta(ct-ct_{3}) \delta(ct_{1}-ct_{2}) 
\t{\delta}^{\mu}_{\;\;\mu_{3}} \delta^{3}(x-x_{3}) 
\left(
\t{\delta}_{\mu_{1}\,\mu_{2}} 
\partial^{\mu_{1}}
\delta^{3}(x_{1}-x_{2}) 
\left[ \h{j}^{(in0)\,0}(x), \h{j}^{(in0)}_{\quad\;\; 0}(x_{1}) \right] 
\right)
\right\}
\\
\nonumber
&&+
\left(\frac{1}{i\hbar c^{2}}\right)^{2} \frac{-q}{mc^{2}}
\\
\label{sus03g_04}
&&
\left\{
\delta(ct-ct_{1})
\theta(ct_{1}-ct_{2}) 
\theta(ct_{2}-ct_{3})
\left(
\t{\delta}^{\mu}_{\;\;\mu_{1}}\partial^{\mu_{1}}
\delta^{3}(x-x_{1})
\right)
\left[ \left[ \h{j}^{(in0)\,0}(x), \h{j}^{(in0)}_{\quad \mu_{2}}(x_{2}) \right], 
\h{j}^{(in0)}_{\quad \mu_{3}}(x_{3}) \right]  
\right.
\\
\label{sus03g_05}
&&+
\delta(ct-ct_{1})
\theta(ct_{1}-ct_{3}) 
\theta(ct_{3}-ct_{2})
\left(
\t{\delta}^{\mu}_{\;\;\mu_{1}}
\partial^{\mu_{1}}
\delta^{3}(x-x_{1})
\right)
\left[ \left[ \h{j}^{(in0)\,0}(x), \h{j}^{(in0)}_{\quad \mu_{3}}(x_{3}) \right], 
\h{j}^{(in0)}_{\quad \mu_{2}}(x_{2}) \right]  
\\
\label{sus03g_06}
&&-
\delta(ct-ct_{2})
\theta(ct_{2}-ct_{3}) 
\delta(ct_{3}-ct_{1})
\t{\delta}^{\mu}_{\;\;\mu_{2}}\delta^{3}(x-x_{2})
\left[ \left[ \h{j}^{(in0)\,0}(x), \h{j}^{(in0)}_{\quad \mu_{3}}(x_{3}) \right], 
\h{j}^{(in0)}_{\quad 0}(x_{1}) \right]  
\\
\nonumber
&&+
\left(
-\delta(ct-ct_{2})
\delta(ct_{2}-ct_{1}) 
\theta(ct_{1}-ct_{3})
+
\delta(ct-ct_{2})
\theta(ct_{2}-ct_{1}) 
\delta(ct_{1}-ct_{3})
\right)
\\
\label{sus03g_07}
&&\hspace{16.0em}
\t{\delta}^{\mu}_{\;\;\mu_{2}}\delta^{3}(x-x_{2})
\left[ \left[ \h{j}^{(in0)\,0}(x), \h{j}^{(in0)}_{\quad 0}(x_{1}) \right], 
\h{j}^{(in0)}_{\quad \mu_{3}}(x_{3}) \right]  
\\
\nonumber
&&+
\left(
-\delta(ct-ct_{3})
\delta(ct_{3}-ct_{1}) 
\theta(ct_{1}-ct_{2})
+
\delta(ct-ct_{3})
\theta(ct_{3}-ct_{1}) 
\delta(ct_{1}-ct_{2})
\right)
\\
\label{sus03g_08}
&&\hspace{16.0em}
\t{\delta}^{\mu}_{\;\;\mu_{3}}\delta^{3}(x-x_{3})
\left[ \left[ \h{j}^{(in0)\,0}(x), \h{j}^{(in0)}_{\quad 0}(x_{1}) \right], 
\h{j}^{(in0)}_{\quad \mu_{2}}(x_{2}) \right]  
\\
\label{sus03g_09}
&&-
\delta(ct-ct_{3})
\theta(ct_{3}-ct_{2}) 
\delta(ct_{2}-ct_{1})
\t{\delta}^{\mu}_{\;\;\mu_{3}}\delta^{3}(x-x_{3})
\left[ \left[ \h{j}^{(in0)\,0}(x), \h{j}^{(in0)}_{\quad \mu_{2}}(x_{2}) \right], 
\h{j}^{(in0)}_{\quad 0}(x_{1}) \right]  
\\
\label{sus03g_10}
&&+
\theta(ct-ct_{1}) \delta(ct_{1}-ct_{2})\theta(ct_{2}-ct_{3})
\t{\delta}_{\mu_{1}\,\mu_{2}}
\partial^{\mu_{1}}\left(
\delta^{3}(x_{1}-x_{2})
\left[ \left[ \h{j}^{(in0)\,\mu}(x), \h{j}^{(in0)}_{\quad 0}(x_{1}) \right], 
\h{j}^{(in0)}_{\quad \mu_{3}}(x_{3}) \right]  
\right)
\\
\label{sus03g_11}
&&-
\theta(ct-ct_{2}) \delta(ct_{2}-ct_{3})\delta(ct_{3}-ct_{1})
\t{\delta}_{\mu_{2}\,\mu_{3}}\delta^{3}(x_{2}-x_{3})
\left[ \left[ \h{j}^{(in0)\,\mu}(x), \h{j}^{(in0)}_{\quad 0}(x_{2}) \right], 
\h{j}^{(in0)}_{\quad 0}(x_{1}) \right]  
\\
\label{sus03g_12}
&&+
\theta(ct-ct_{3}) \delta(ct_{3}-ct_{1})\theta(ct_{1}-ct_{2})
\t{\delta}_{\mu_{3}\,\mu_{1}}
\left( \partial^{\mu_{1}}
\delta^{3}(x_{3}-x_{1})
\right)
\left[ \left[ \h{j}^{(in0)\,\mu}(x), \h{j}^{(in0)}_{\quad 0}(x_{3}) \right], 
\h{j}^{(in0)}_{\quad \mu_{2}}(x_{2}) \right]  
\\
\nonumber
&&+
\left(
-\delta(ct-ct_{1}) \theta(ct_{1}-ct_{2})\delta(ct_{2}-ct_{3})
+\theta(ct-ct_{1}) \delta(ct_{1}-ct_{2})\delta(ct_{2}-ct_{3})
\right)
\\
\label{sus03g_13}
&&\hspace{16.0em}
\t{\delta}_{\mu_{2}\,\mu_{3}}\delta^{3}(x_{2}-x_{3})
\left[ \left[ \h{j}^{(in0)\,\mu}(x), \h{j}^{(in0)}_{\quad 0}(x_{1}) \right], 
\h{j}^{(in0)}_{\quad 0}(x_{2}) \right]
\\
\label{sus03g_14}
&&+
\theta(ct-ct_{2}) \theta(ct_{2}-ct_{3})\delta(ct_{3}-ct_{1})
\left(
\t{\delta}_{\mu_{3}\,\mu_{1}}
 \partial^{\mu_{1}}
\delta^{3}(x_{3}-x_{1})
\right)
\left[ \left[ \h{j}^{(in0)\,\mu}(x), \h{j}^{(in0)}_{\quad \mu_{2}}(x_{2}) \right], 
\h{j}^{(in0)}_{\quad 0}(x_{3}) \right]
\\
\label{sus03g_15}
&&\left.+
\theta(ct-ct_{3}) \theta(ct_{3}-ct_{1})\delta(ct_{1}-ct_{2})
\t{\delta}_{\mu_{1}\,\mu_{2}}
\partial^{\mu_{1}} \left(
\delta^{3}(x_{1}-x_{2})
\left[ \left[ \h{j}^{(in0)\,\mu}(x), \h{j}^{(in0)}_{\quad \mu_{3}}(x_{3}) \right], 
\h{j}^{(in0)}_{\quad 0}(x_{1}) \right]
\right)
\right\}
\\
\nonumber
&&+
\left(\frac{1}{i\hbar c^{2} }\right)^{3}
\\
\nonumber
&&
\left\{ 
\left(
-\delta(ct-ct_{1}) \theta(ct_{1}-ct_{2})  \theta(ct_{2}-ct_{3})
+\theta(ct-ct_{1}) \delta(ct_{1}-ct_{2})  \theta(ct_{2}-ct_{3})
\right)\right.
\\
\label{sus03g_16}
&&\hspace{16.0em}
\left[ \left[ \left[ \h{j}^{(in0)\,\mu}(x), \h{j}^{(in0)}_{\quad 0}(x_{1}) \right], 
\h{j}^{(in0)}_{\quad \mu_{2}}(x_{2}) \right] ,
\h{j}^{(in0)}_{\quad \mu_{3}}(x_{3}) \right]
\\
\nonumber
&&+
\left(
-\delta(ct-ct_{1}) \theta(ct_{1}-ct_{3})  \theta(ct_{3}-ct_{2})
+\theta(ct-ct_{1}) \delta(ct_{1}-ct_{3})  \theta(ct_{3}-ct_{2})
\right)
\\
\label{sus03g_17}
&&\hspace{16.0em}
\left[ \left[ \left[ \h{j}^{(in0)\,\mu}(x), \h{j}^{(in0)}_{\quad 0}(x_{1}) \right], 
\h{j}^{(in0)}_{\quad \mu_{3}}(x_{3}) \right] ,
\h{j}^{(in0)}_{\quad \mu_{2}}(x_{2}) \right]
\\
\label{sus03g_18}
&&-
\theta(ct-ct_{2}) \theta(ct_{2}-ct_{3})  \delta(ct_{3}-ct_{1})
\left[ \left[ \left[ \h{j}^{(in0)\,\mu}(x), \h{j}^{(in0)}_{\quad \mu_{2}}(x_{2}) \right], 
\h{j}^{(in0)}_{\quad \mu_{3}}(x_{3}) \right] ,
\h{j}^{(in0)}_{\quad 0}(x_{1}) \right]
\\
\nonumber
&&+
\left(
-\theta(ct-ct_{2}) \delta(ct_{2}-ct_{1})  \theta(ct_{1}-ct_{3})
+\theta(ct-ct_{2}) \theta(ct_{2}-ct_{1})  \delta(ct_{1}-ct_{3})
\right)
\\
\label{sus03g_19}
&&\hspace{16.0em}
\left[ \left[ \left[ \h{j}^{(in0)\,\mu}(x), \h{j}^{(in0)}_{\quad \mu_{2}}(x_{2}) \right], 
\h{j}^{(in0)}_{\quad 0}(x_{1}) \right] ,
\h{j}^{(in0)}_{\quad \mu_{3}}(x_{3}) \right]
\\
\nonumber
&&+
\left(
-\theta(ct-ct_{3}) \delta(ct_{3}-ct_{1})  \theta(ct_{1}-ct_{2})
+\theta(ct-ct_{3}) \theta(ct_{3}-ct_{1})  \delta(ct_{1}-ct_{2})
\right)
\\
\label{sus03g_20}
&&\hspace{16.0em}
\left[ \left[ \left[ \h{j}^{(in0)\,\mu}(x), \h{j}^{(in0)}_{\quad \mu_{3}}(x_{3}) \right], 
\h{j}^{(in0)}_{\quad 0}(x_{1}) \right] ,
\h{j}^{(in0)}_{\quad \mu_{2}}(x_{2}) \right]
\\
\label{sus03g_21}
&&\left.-
\theta(ct-ct_{3}) \theta(ct_{3}-ct_{2})  \delta(ct_{2}-ct_{1})
\left[ \left[ \left[ \h{j}^{(in0)\,\mu}(x), \h{j}^{(in0)}_{\quad \mu_{3}}(x_{3}) \right], 
\h{j}^{(in0)}_{\quad \mu_{2}}(x_{2}) \right] ,
\h{j}^{(in0)}_{\quad 0}(x_{1}) \right]
\right\}.
\eeqna
\end{subequations}
\end{footnotesize}
In the following, we prove the next equation, which leads to the gauge invariance.
\beqna
\label{sus03g_1stgroup}
\mbox{(\ref{sus03g_16}) + (\ref{sus03g_04}) +(\ref{sus03g_10}) +(\ref{sus03g_19})
+(\ref{sus03g_18})+(\ref{sus03g_14}) } &=&0,
\\
\label{sus03g_2ndgroup}
\mbox{(\ref{sus03g_17}) + (\ref{sus03g_05}) +(\ref{sus03g_12}) +(\ref{sus03g_20})
+(\ref{sus03g_21})+(\ref{sus03g_15}) } &=&0,
\\
\label{sus03g_3rdgroup}
\mbox{(\ref{sus03g_13}) + (\ref{sus03g_01}) +(\ref{sus03g_11}) } &=&0,
\\
\label{sus03g_4thgroup}
\mbox{(\ref{sus03g_09}) + (\ref{sus03g_03}) +(\ref{sus03g_08}) } &=&0,
\\
\label{sus03g_5thgroup}
\mbox{(\ref{sus03g_06}) + (\ref{sus03g_02}) +(\ref{sus03g_07}) } &=&0.
\eeqna
%
%
\begin{enumerate}
\item \underline{\eq(sus03g_1stgroup):}\;
The first term of (\ref{sus03g_16}) offsets  (\ref{sus03g_04}), using \eq(j0jmu).
To the inner double commutator in the second term of (\ref{sus03g_16}),
the nest Jacobi identity is applied:
\beqna
\label{jacobi16}
&&\left[ \left[ \left[ \h{j}^{(in0)\,\mu}(x), \h{j}^{(in0)}_{\quad 0}(x_{1}) \right], 
\h{j}^{(in0)}_{\quad \mu_{2}}(x_{2}) \right] ,
\h{j}^{(in0)}_{\quad \mu_{3}}(x_{3}) \right]
\\
\nonumber
&=&-\left[ \left[ \left[ \h{j}^{(in0)}_{\quad 0}(x_{1}),\h{j}^{(in0)}_{\quad \mu_{2}}(x_{2}) \right], 
\h{j}^{(in0)\,\mu}(x) \right] ,
\h{j}^{(in0)}_{\quad \mu_{3}}(x_{3}) \right]
\\
\nonumber
&&
-\left[ \left[ \left[ \h{j}^{(in0)}_{\quad \mu_{2}}(x_{2}) , \h{j}^{(in0)\,\mu}(x) \right], 
\h{j}^{(in0)}_{\quad 0}(x_{1})\right] ,
\h{j}^{(in0)}_{\quad \mu_{3}}(x_{3}) \right].
\eeqna
Furthermore, the inner commutator (, assuming at the simultaneous time) in the first term of  \eq(jacobi16) ,
one may apply \eq(j0jmu).
The part including this factor in the second term of (\ref{sus03g_16}) offsets the term 
(\ref{sus03g_10}).
In the second term of (\ref{sus03g_16}), the part including the second term of \eq(jacobi16)
offsets the first term of (\ref{sus03g_19}).
Up to now, 
(\ref{sus03g_16})+(\ref{sus03g_04})+(\ref{sus03g_10})+ the first term of (\ref{sus03g_19})$=0$
has been shown.

Next, to the outer double commutator in the second term of (\ref{sus03g_18}),
let us apply the next Jacobi identity,
\beqna
\label{jacobi18}
&&\left[ \left[ \left[ \h{j}^{(in0)\,\mu}(x), \h{j}^{(in0)}_{\quad \mu_{2}}(x_{2}) \right], 
\h{j}^{(in0)}_{\quad \mu_{3}}(x_{3}) \right] ,
\h{j}^{(in0)}_{\quad 0}(x_{1}) \right]
\\
\nonumber
&=&-\left[ \left[ \h{j}^{(in0)}_{\quad \mu_{3}}(x_{3}) ,\h{j}^{(in0)}_{\quad 0}(x_{1}) \right] ,
\left[ \h{j}^{(in0)\,\mu}(x), \h{j}^{(in0)}_{\quad \mu_{2}}(x_{2}) \right]\right]
\\
\nonumber
&&+\left[ \left[ \left[ \h{j}^{(in0)\,\mu}(x), \h{j}^{(in0)}_{\quad \mu_{2}}(x_{2}) \right], 
\h{j}^{(in0)}_{\quad 0}(x_{1})\right] ,
\h{j}^{(in0)}_{\quad \mu_{3}}(x_{3}) 
 \right].
\eeqna
The commutator in the first term of \eq(jacobi18):
$\left[ \h{j}^{(in0)}_{\quad \mu_{3}}(x_{3}) ,\h{j}^{(in0)}_{\quad 0}(x_{1}) \right] $
is the commutator at the simultaneous time, therefore, we may use \eq(j0jmu).
The part including this factor in (\ref{sus03g_18}) offsets the term 
(\ref{sus03g_14}).
In (\ref{sus03g_18}), the part including the second term of \eq(jacobi18)
offsets the second term of (\ref{sus03g_19}).
Up to now, 
(\ref{sus03g_18})+(\ref{sus03g_14})+ the second term of (\ref{sus03g_19})$=0$
is shown.

Together with the previous result, \eq(sus03g_1stgroup) holds.
\item \underline{\eq(sus03g_2ndgroup):}\;
This equation is \eq(sus03g_1stgroup) with the replacement
$x_{2} \leftrightarrow x_{3}$ and $\mu_{2} \leftrightarrow \mu_{3}$, therefore,
\eq(sus03g_2ndgroup) holds.
 \item \underline{\eq(sus03g_3rdgroup):}\;
 The first term of (\ref{sus03g_13}) offsets (\ref{sus03g_01}), using \eq(j0jmu).
 
 To the double commutator in the second term of (\ref{sus03g_13}),
the nest Jacobi identity is applied:
\beqna
\label{jacobi13}
&&
\left[ \left[ \h{j}^{(in0)\,\mu}(x), \h{j}^{(in0)}_{\quad 0}(x_{1}) \right], 
\h{j}^{(in0)}_{\quad 0}(x_{2}) \right]
\\
\nonumber
&=&
-\left[ \left[  \h{j}^{(in0)}_{\quad 0}(x_{1}),\h{j}^{(in0)}_{\quad 0}(x_{2}) \right], 
 \h{j}^{(in0)\,\mu}(x)\right]
-\left[ \left[ \h{j}^{(in0)}_{\quad 0}(x_{2}), \h{j}^{(in0)\,\mu}(x) \right], 
\h{j}^{(in0)}_{\quad 0}(x_{1}) \right]
\\
\nonumber
&=&
\left[ \left[  \h{j}^{(in0)\,\mu}(x),\h{j}^{(in0)}_{\quad 0}(x_{2}) \right], 
\h{j}^{(in0)}_{\quad 0}(x_{1}) \right]
\eeqna
In the above,  we use the inner commutator (at the simulatenesou time) in the first term of 
the second hand side becomes zero, using \eq(j0jmu).
The second term of (\ref{sus03g_13}) includes the factor of \eq(jacobi13)
and offsets (\ref{sus03g_11}).

As a result, \eq(sus03g_3rdgroup) holds.
\item \underline{\eq(sus03g_4thgroup):}\;
To the double commutator in (\ref{sus03g_09}), we apply the next Jacobi identity:
\beqna
\label{jacobi09}
&&
\left[ \left[ \h{j}^{(in0)\,0}(x), \h{j}^{(in0)}_{\quad \mu_{2}}(x_{2}) \right], 
\h{j}^{(in0)}_{\quad 0}(x_{1}) \right]  
\\
\nonumber
&=&
-\left[ \left[ \h{j}^{(in0)}_{\quad \mu_{2}}(x_{2}),\h{j}^{(in0)}_{\quad 0}(x_{1}) \right], 
\h{j}^{(in0)\,0}(x) \right]  
-\left[ \left[ \h{j}^{(in0)}_{\quad 0}(x_{1}) , \h{j}^{(in0)\,0}(x) \right], 
\h{j}^{(in0)}_{\quad \mu_{2}}(x_{2})
\right].  
\eeqna

The inner commutator in the first term of \eq(jacobi09)
is the commutator at the simultaneous time, therefore, we may use \eq(j0jmu).
The part including this factor in (\ref{sus03g_09}) offsets the term 
(\ref{sus03g_03}).
In (\ref{sus03g_09}), the part including the second term of \eq(jacobi09)
offsets the second term of (\ref{sus03g_08}).
The first term of (\ref{sus03g_08}) is zero, because the inner commutator
included in this term is commutator at the simultaneous time and leads to zero, using \eq(j0jmu).

Therefore, \eq(sus03g_4thgroup) holds.
\item \underline{\eq(sus03g_5thgroup):}\;
This equation is \eq(sus03g_4thgroup) with the replacement
$x_{2} \leftrightarrow x_{3}$ and $\mu_{2} \leftrightarrow \mu_{3}$, therefore,
\eq(sus03g_5thgroup) holds.
\end{enumerate}
As the summery, \eqss(sus03g_1stgroup)-(sus03g_5thgroup) hold and
the third order nonlinear single susceptibility Eq.(42) in the main text maintains the gauge invariance
Eq.(25) in the main text.